\documentclass[twocolumn,reprint,superscriptaddress,amsmath,amssymb,aps,pra]{revtex4-2}
\usepackage{paracol}
\usepackage{units}
\usepackage{graphicx,subfigure}
\usepackage{mathrsfs}
\usepackage{bm}% bold math
\usepackage{textcomp}
\usepackage{url}
\usepackage{dsfont}
\usepackage{braket}
\usepackage{xcolor}
\usepackage[compact]{titlesec}
\usepackage[colorlinks=true,linkcolor=magenta,citecolor=blue]{hyperref}%
\usepackage{todonotes}
\usepackage{empheq}
\usepackage{multirow}
\usepackage{titlesec}
\usepackage{cleveref}
\definecolor{mypink1}{rgb}{0.858, 0.188, 0.478}

\usepackage{cancel}
\usepackage{float}

\titleformat{\subsubsection}
 {\normalfont\normalsize\itshape}{\thesubsubsection}{1em}{}
\titlespacing*{\subsubsection}{0pt}{3.25ex plus 1ex minus.2ex}{0ex plus.2ex}
\begin{document} 
\title{Non-Hermitian Second-Order Topological Phases and Bipolar Skin Effect in Photonic Kagome Crystals}

\author{Xiaosen Yang} \altaffiliation{yangxs@ujs.edu.cn} \affiliation{Department of Physics, Jiangsu University, Zhenjiang, 212013, China}

\author{Yaru Feng} \affiliation{Department of Physics, Jiangsu University, Zhenjiang, 212013, China}

\author{Abdul Wahab} \affiliation{Department of Physics, Jiangsu University, Zhenjiang, 212013, China}

\author{Hao Geng} \affiliation{Department of Physics, Jiangsu University, Zhenjiang, 212013, China}

\date{\today}

\begin{abstract}
Non-Hermitian photonics provides a fertile platform for exploring phenomena with no Hermitian counterparts, including the non-Hermitian skin effect and exceptional points, with direct relevance for integrated photonic technologies. In this work, we investigate the properties of non-Hermitian second-order topological phases by constructing a photonic kagome crystal with balanced gain and loss, and reveal the interplay between higher-order topology and the non-Hermitian skin effect. We demonstrate that non-Hermiticity not only lifts the degeneracy of the topological corner modes but also drives bulk states to accumulate at corners, giving rise to bipolar non-Hermitian skin effect. By defining the point-gap topology, we uncover the fundamental topological origin of the non-Hermitian skin effect. More interestingly, the non-Hermitian skin effect induces a fundamental breakdown of the conventional bulk-boundary correspondence based on the Bloch band theory. Our findings establish a general framework for non-Hermitian higher-order photonic systems and open avenues toward tailorable topological photonic devices exploiting non-Hermitian enhanced localization.

\end{abstract}

\maketitle
%%%%%%%%%%%%%%%%%%%%%%%%%%%%%%%%%%%%%%%%% 
\section{Introduction}

The exploration of non-Hermitian physics~\cite{PhysRevLett.80.5243,PhysRevLett.115.040402,PhysRevLett.118.040401,PhysRevLett.123.206404,PhysRevLett.123.246801,PhysRevB.99.201103,Ashida02072020,RevModPhys.93.015005,PhysRevB.105.165137,PhysRevB.106.115107,PhysRevB.110.014106,PhysRevB.109.064203,PhysRevB.109.125420,PhysRevB.111.075132,PhysRevB.111.094109,PhysRevB.111.195419} has revealed a wealth of interesting phenomena entirely absent in Hermitian systems~\cite{yao2018edge,PhysRevLett.123.170401,PhysRevResearch.1.023013,PhysRevLett.125.126402,li2020critical,PhysRevLett.124.086801,PhysRevLett.125.118001,PhysRevB.102.205118,zhang2021observation,zhang2022universal,PhysRevB.106.195425,HU202551}, among which the non-Hermitian skin effect (NHSE)~\cite{PhysRevB.102.241202,PhysRevB.104.165117,gu2022transient,PhysRevLett.129.070401,lin2023topological,PhysRevX.13.021007,li2024observation} has attracted particular attention in condensed matter physics~\cite{el2018non,PhysRevLett.121.136802,PhysRevLett.123.066404,pan2020non,Song:22}, cold atomic physics~\cite{PhysRevLett.124.250402,PhysRevA.109.053311,PhysRevA.110.063320,PhysRevA.109.053305}, photonics~\cite{doi:10.1126/science.aay1064, Longhi:19,Weidemann2020topological,li2020higher,li2021non, Wang:23, PhysRevB.108.035406,reisenbauer2024non, 10.1063/5.0183826,Zheng2024,Roccati2024,PhysRevB.110.094308,PhysRevB.110.104103,Feng2025} and acoustics~\cite{PhysRevLett.122.195501,rosendo2019multiple,gao2021non,PhysRevApplied.16.057001,PhysRevLett.131.066601,hu2024engineering,huang2024acoustic}. Unlike in Hermitian systems, where bulk states are extended, the bulk states of non-Hermitian systems can exhibit a macroscopic accumulation at the boundaries~\cite{PhysRevLett.124.086801}, profoundly altering spectral and transport properties
~\cite{PhysRevB.107.125155,PhysRevB.110.045138,z9m1-3mwb}.
This NHSE challenges the conventional bulk-boundary correspondence based on the Bloch band theory \cite{PhysRevLett.125.226402,xiao2020non,helbig2020generalized,PhysRevB.103.075126,xiao2024restoration,Ji_2024} and introduces new opportunities to manipulate light on photonic platforms.

Photonic crystals, whose geometric structure and material parameters can be easily tuned
~\cite{JOANNOPOULOS1997165,Akahane2003,Cubukcu2003,he2020quadrupole,ma2021deep}, have already been proven to be a good platform for performing topological phenomena~\cite{ota2018topological,doi:10.1126/science.aar4005,yang2019realization,PhysRevLett.122.233902,lu2014topological} such as higher-order topological insulators (HOTIs)~\cite{lu2014topological,khanikaev2017two, Xie:18,RevModPhys.91.015006,OtaTakataOzawaAmoJiaKanteNotomiArakawaIwamoto+2020+547+567, kim2020recent,xue2021topological,XIE2023255,ZHANG2024111353}. In contrast to ﬁrst-order topological phases, an $n$th-order topological phase in $d$ spatial dimension hosts $(d-n)$-dimensional robust topological corner/hinge modes ($1<n\leq d$)~\cite{PhysRevB.97.205136,zhang2019dimensional,xie2021higher}. Both theoretical and experimental studies have confirmed that long-range interactions in photonic crystals can result in the formation of a new class of higher-order topological (HOT) corner modes~\cite{li2020higher}. More importantly, photonic crystals serve as a good platform for investigating non-Hermitian physics. Non-Hermitian topological phenomena in photonic crystals can be achieved by incorporating gain and loss mechanisms~\cite{Wang:23,PhysRevB.108.035406}. The non-Hermiticity significantly breaks the degeneracy of topological corner modes~\cite{Wang:23}. The NHSE as a novel mechanism for light localization has also recently been explored in photonic crystals. The existence of photonic corner skin modes has been confirmed in two-dimensional non-Hermitian photonic crystals~\cite{PhysRevB.108.035406}. However, the interplay between NHSE and higher-order topology in photonic crystals remains unclear and needs to be uncovered.

In this work, we design a non-Hermitian photonic kagome crystal with balanced gain and loss that realizes HOT phases and the NHSE. Our results demonstrate that non-Hermiticity not only lifts the degeneracy of topological corner and edge modes but also drives bulk states to accumulate at opposite corners, giving rise to a bipolar NHSE. By defining a winding number to characterize the point-gap topology of the complex eigenfrequency spectrum, we reveal the topological origin of the bipolar NHSE and clarify its role in reshaping HOT phase transitions. Crucially, the bipolar NHSE leads to a breakdown of the conventional Bloch bulk-boundary correspondence: bulk topological invariants defined on the Brillouin zone no longer provide a complete characterization of the emergence of robust topological edge and corner modes.

\section{Non-Hermitian photonic kagome crystal}\label{Non-Hermitian Photonic}
\begin{figure}
\includegraphics[width=0.48\textwidth]{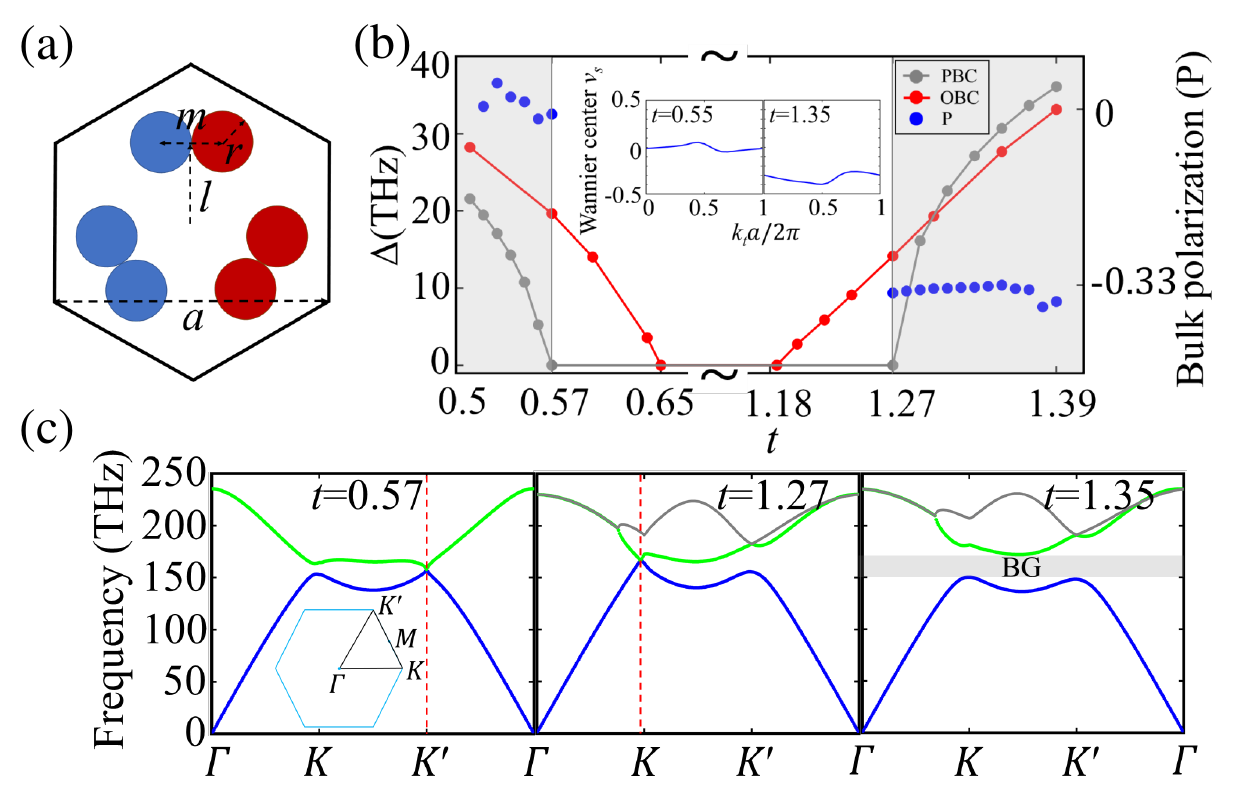}
\caption{\textbf{Topological phase transition and band structure of non-Hermitian photonic kagome crystals.} (a) Unit cell of the gyromagnetic non-Hermitian photonic kagome crystal, where gain and loss are introduced in red and blue dielectric columns, respectively. (b) Photonic band gap as functions of $ t $ at $\gamma=6$ under PBC and OBC. The blue circular dots represent the Bloch topological bulk polarization. The inset shows the Wannier center at $ t=0.55 $ and $ t=1.35 $, respectively. (c) The band diagrams for $t=0.57$, $1.27$ and $1.35$ with $\gamma=6$. 
}\label{figpbc} 
\end{figure}

To investigate the interplay between topology and the NHSE, we design a non-Hermitian gyromagnetic photonic crystal characterized by nontrivial bulk polarization. All numerical simulations are performed using the commercial finite-element package COMSOL Multiphysics. Specifically, we implement balanced gain (red) and loss (blue) in six dielectric columns of a photonic kagome crystal by modifying their relative permittivity to $\varepsilon_r = 12 \pm i\gamma$, where $\gamma$ quantifies the non-Hermitian strength, as illustrated in Fig.~\ref{figpbc}(a). In addition, time-reversal symmetry is explicitly broken by applying an external magnetic field along the $z$ axis, under which the relative permeability tensor takes the form~\cite{PhysRevB.108.035406}:
%%%%%%%%%%%%%%%%%%%%%%%%%%%%%%%%%%
\begin{align}
\mathbf{\mu_r}=\left[ \begin{matrix}
\mu&		i\kappa&		0\\
-i\kappa&		\mu&		0\\
0&		0&		\mu_{0}\\
\end{matrix} \right], 
\end{align}
%%%%%%%%%%%%%%%%%%%%%%%%%%%%%%%%%%
with $\kappa= 2.16 \mu_{0}$ and $\mu= 2.2 \mu_{0}$, where $\mu_{0}$ is the permeability of the vacuum. The complex photonic kagome crystal features a lattice constant of $a=0.69 ~\mu m$, with dielectric cylinders of radius $r=a/12$, and the spacing between the dimer columns set to $m=2r$.
The normal distance between the dimer center and the lattice center is $b = \sqrt{3}a/6$, the distance $l$ from the center of the lattice to the center of the dimer is given by $l=t\cdot b$~\cite{li2020higher,Wang:23}, and the complex unit cell can be shrunken or expanded by multiplying the scaling parameter $t$. 

%%%%%%%%%%%%%%%%%

Without loss of generality, we set $\gamma = 6$ to illustrate the distinctive properties of the non-Hermitian photonic crystal. The photonic band gap of the system as a function of $t$ under periodic-boundary conditions (PBC) and open-boundary conditions (OBC) are depicted in Fig.~\ref{figpbc}(b), with red and gray line chart representing phase boundaries under OBC and PBC, respectively. The photonic band gaps under PBC and OBC are different in essence. Under PBC, the photonic band gap closes at $t = 0.57$ and reopens at $t = 1.27$, where a Dirac point appears at the $K$ ($K'$) point as shown in Fig.~\ref{figpbc}(c). To focus on the topological phase transition associated with this gap, we consider only the two lowest bands (blue and green), neglecting the higher energy band (gray). As $t$ increases from $0.57$ to $1.27$, the Dirac point moves from $K'$ to $K$ in the Brillouin zone, indicating a stable gapless band structure in this region. There is a photonic band gap when $t > 1.27$ as shown in the right panel of Fig.~\ref{figpbc}(c), with the gray shaded region marking the photonic band gap. According to the conventional bulk-boundary correspondence based on Bloch band theory, the topological properties of the gapped photonic crystal can be characterized by the bulk polarization for Wannier-type HOTIs~\cite{atala2013direct}.  The bulk polarization is associated with the Wannier centers and can be defined as 
~\cite{PhysRevA.100.032102,li2020higher}:
\begin{equation}
P_s = \frac{1}{N_k} \sum_{j,k_t} \nu^j_s(k_t).
\tag{2}
\label{eq:formula2}
\end{equation}
The Brillouin zone is discretized along the reciprocal lattice vectors $\mathbf{b}_1$ and $\mathbf{b}_2$, and the discrete momentum are labeled as $(k_t, k_s)$, where $k_t$ and $k_s$ denote sampling points along $\mathbf{b}_1$ and $\mathbf{b}_2$, respectively.
$N_k$ is the number of unit cells along $\mathbf{b}_1$ direction. $j$ represents the index of the occupied band. The Wilson loop is defined as~\cite{he2020quadrupole,Wang_2019,PhysRevA.104.033501}:
%\begin{equation}
    % W_{k_s \rightarrow 2\pi + k_s, k_t} = \frac{1}{2\pi} \operatorname{Im} \left[ \ln \prod_{k_s=0}^{2\pi-\delta k} U_{\mathbf{k}_{(k_s,k_t)} \to \mathbf{k}_{(k_s+\delta k,k_t)}} \right],
    %\tag{3}
   % \label{eq:formula4}
%\end{equation}  
\begin{equation}
 W_{k_s \rightarrow 2\pi + k_s, k_t} = \prod_{k_s=0}^{2\pi-\delta k} U_{\mathbf{k}_{(k_s,k_t)} \to \mathbf{k}_{(k_s+\delta k,k_t)}}, 
    \tag{3}
    \label{eq:formula3}
\end{equation}
with
\begin{equation}
     U_{\mathbf{k}_\alpha \to \mathbf{k}_\beta} = \frac{{\langle u_{\mathbf{k}_\alpha} (\mathbf{r})} | u_{\mathbf{k}_\beta}(\mathbf{r}) \rangle}
     {|\langle u_{\mathbf{k}_\alpha}(\mathbf{r}) | u_{\mathbf{k}_\beta} (\mathbf{r})\rangle|}.
\tag{4}
\label{eq:formula4}
\end{equation}
%\begin{equation}
%U_{s,\mathbf{k}}^{mn} = \langle u_{\mathbf{k}}^{m} | \varepsilon | u_{\mathbf{k}+\Delta_{s}}^{n} \rangle
%\tag{4}
%\label{eq:formula5}
%\end{equation}
The Wannier center $\nu^j_s(k_t)$ are encoded in the phases of the Wilson loop eigenvalues
\begin{equation}
W_{s,\mathbf{k}}\ket{v_{s,\mathbf{k}}^{j}} = e^{i2\pi\nu_{s}^{j}(k_{t})}\ket{v_{s,\mathbf{k}}^{j}}.
\tag{5}
\label{eq:formula5}
\end{equation}

$u_{\mathbf{k}}(\mathbf{r})$ is cell-periodic electric field with $u_{\mathbf{k}}(\mathbf{r}) = u_{\mathbf{k}}(\mathbf{r} + \mathbf{R})$ for any lattice vector $\mathbf{R}$. The momentum $k_t, k_s\in [0, \delta k, \frac{4\pi}{\sqrt{3}a})$ and $\delta k = \frac{1}{N_k} \frac{4\pi}{\sqrt{3}a}$.

Accordingly, we calculate the Wilson loop along $\mathbf{b}_2$ for the lowest energy band, and the bulk polarization $P_s$ is denoted simply as $P$ throughout this paper. Figure~\ref{figpbc}(b) shows the bulk polarization of the lowest Wannier band as a function of $t$ (blue dots), with the Wannier centers at $t=0.55$ and $t=1.35$ illustrated in the inset. The bulk polarization is $0$ for $t<0.57$ and quantized to $-1/3$ for $t>1.27$, indicating topologically trivial and nontrivial phases, respectively, according to the Bloch bulk-boundary correspondence. In contrast, a finite photonic band gap appears at $t=0.57$ and $t=1.27$ under OBC, signaling the breakdown of Bloch band theory in non-Hermitian systems.
% In contrast, a finite photonic band gap appears at the PBC gap-closing points $t=0.57$ and $t=1.27$ under OBC, signaling the breakdown of Bloch band theory in non-Hermitian systems. 
In the following, we further demonstrate the failure of the conventional bulk-boundary correspondence: the bulk polarization $P$, defined on the Brillouin zone, fails to characterize the emergence of topological corner modes, a hallmark of topological phases in Hermitian systems.

\begin{figure*}[t]
\centering
\includegraphics[width=0.9\textwidth]{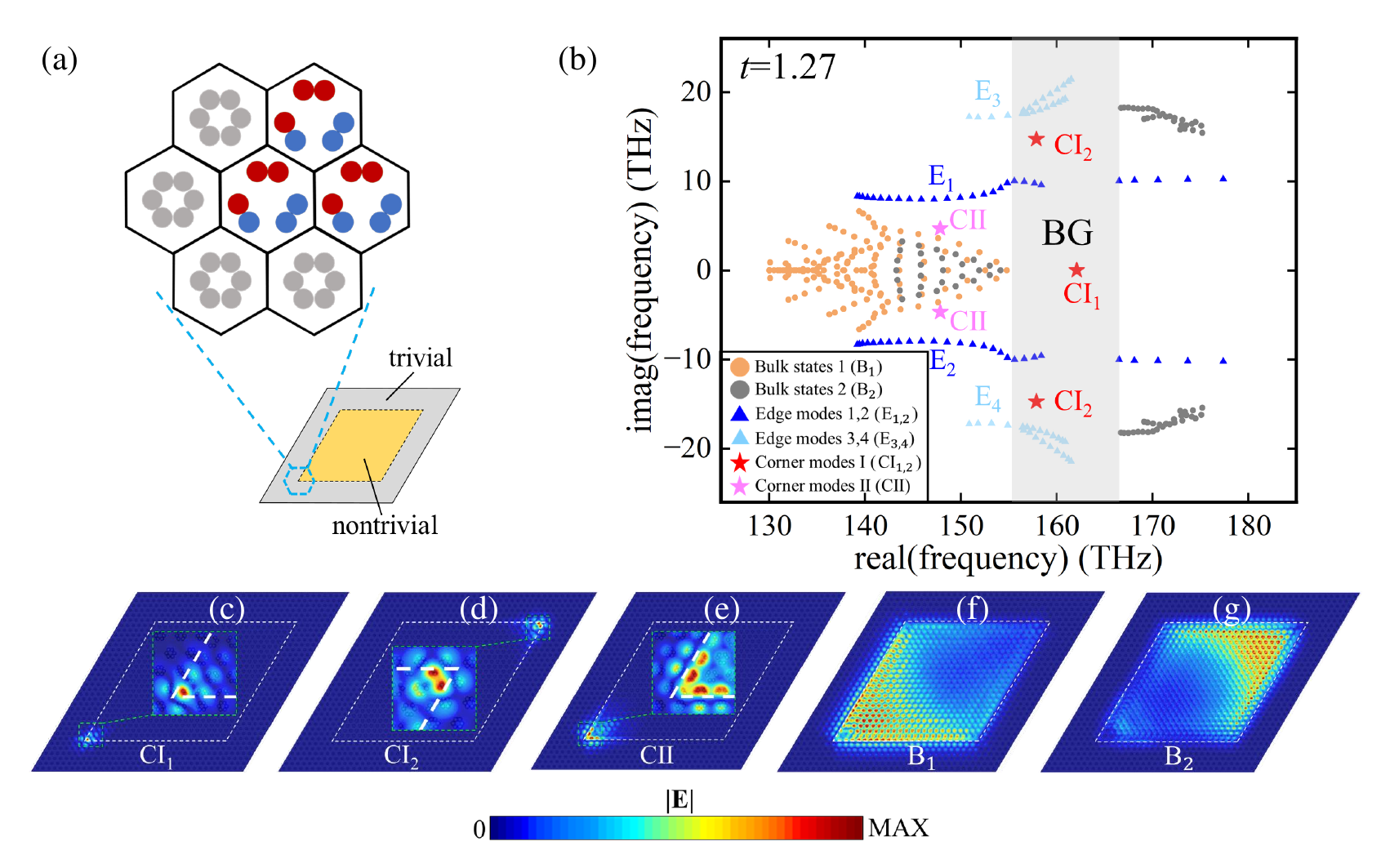}
\caption
{\textbf{NHSE and HOT corner modes in non-Hermitian photonic crystal rhombic supercell ($t=1.27$) under OBC.}
(a) Rhombic supercell composed of $24\times24$ topologically nontrivial non-Hermitian unit cells, enclosed by six layers of trivial Hermitian unit cells.
(b) Complex eigenfrequencies spectrum showing bulk (B$_{1}$, B$_{2}$), edge (E$_{1}$-E$_{4}$), and topological corner modes (CI and CII). The gray-shaded region denotes the photonic band gap (BG). CI$_1$ has a purely real eigenfrequency, while CI$_2$ and CII appear as complex-conjugate pairs.
(c), (d) Normalized eigenmode field profiles of the type-I HOT corner modes CI$_1$ and CI$_2$.
(e) Normalized eigenmode field profile of the type-II HOT corner mode CII.
(f), (g) Normalized eigenmode field profiles of bulk states B$_1$ and B$_2$, exhibiting the NHSE.}
\label{figobc}
\end{figure*}
%%%%%%%%%%%%%%%%%%%%%

\section{HOT corner modes and breakdown of Bloch bulk-boundary correspondence}\label{Unconventional bulk-boundary correspondence}

To investigate the topological properties of the non-Hermitian photonic crystal, we construct a rhombic supercell based on the extended kagome lattice, as shown in Fig.~\ref{figobc}(a). 
A comparison with the triangular supercell (see Appendix) shows that the rhombic geometry exhibits a more pronounced non-Hermitian skin effect, leading to a more concentrated distribution of edge modes in the complex eigenfrequency spectrum. We therefore focus on the rhombic supercell, which provides a clearer platform for elucidating the underlying topological phenomena.
The supercell consists of $24\times24$ non-Hermitian unit cells, surrounded by six layers of Hermitian unit cells that define domain walls in the center and at the outer edges. In the Hermitian cells, the cylinder radius is fixed at $a/12$, and the dimer displacement from the unit cell center is $0.57b$. Figure~\ref{figobc}(b) shows the complex eigenfrequencies spectrum of the non-Hermitian kagome supercell under OBC at $t=1.27$, where the PBC band gap closes. A large photonic band gap appears between 154.8 THz and 166.7 THz, within which two types of HOT corner modes are identified: type-I corner modes (CI$_1$, CI$_2$, red stars) and type-II corner modes (CII, pink stars). In the Hermitian case, both types of corner modes are triply degenerate due to generalized chiral symmetry \cite{li2020higher}. Non-Hermiticity splits the triply degenerate type-I corner modes into two categories: a single corner mode with a purely real eigenfrequency (CI$_1$), localized at the lower-left corner [Fig.~\ref{figobc}(c)], and a pair of complex-conjugate modes (CI$_2$), localized at the upper-right corner [Fig.~\ref{figobc}(d)]. Thus, non-Hermiticity lifts the degeneracy not only in the eigenfrequencies but also in the corresponding eigenmode field profiles of the type-I corner modes.
In Hermitian photonic crystals, long-range interactions drive the localization of topological edge modes at the corners, giving rise to type-II HOT corner modes~\cite{li2020higher}.  
In the presence of non-Hermiticity, these type-II corner modes evolve into paired corner modes with complex-conjugate eigenfrequencies (CII), localized exclusively at the lower-left corner [Fig.~\ref{figobc}(e)].

The emergence and evolution of these corner modes provide clear evidence for the breakdown of the Bloch bulk-boundary correspondence in non-Hermitian photonic crystals. The failure of the Bloch bulk-boundary correspondence is induced by the NHSE in our non-Hermitian photonic crystal. 
Figures~\ref{figobc}(f) and~\ref{figobc}(g) present the normalized eigenmode field profiles of bulk states B$_1$ and B$_2$ (orange and gray dots in Fig.~\ref{figobc}(b)), which exhibit pronounced bipolar NHSE: all the bulk states in B$_1$ are localized at the lower-left corner, while those in B$_2$ are localized at the upper-right corner of the supercell.

\begin{figure}[t]
\includegraphics[width=0.48\textwidth]{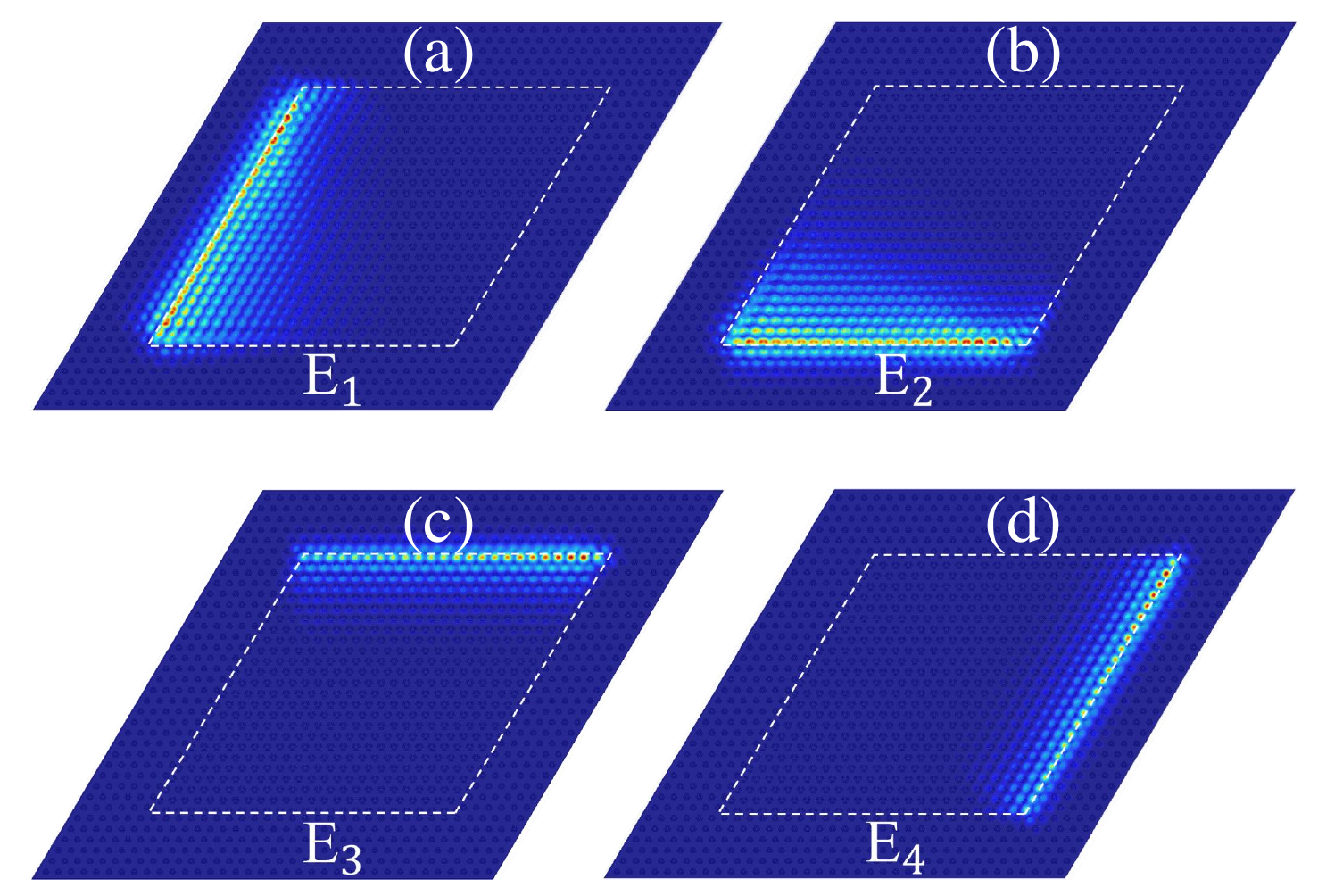}
\caption{\textbf{Topological edge modes in non-Hermitian photonic crystal rhombic supercell ($t=1.27$) under OBC.} (a)-(d) Normalized eigenmode field profiles corresponding to topological edge modes E$_{1}$-E$_{4}$ in Fig.~\ref{figobc}(b). These edge modes localized at four different edges of the supercell, respectively.
}\label{figedgestates}
\end{figure}

%%%%%%%%%%%%%%%%%%%%%%%%%%%
\begin{figure}[b]
\includegraphics[width=0.48\textwidth]{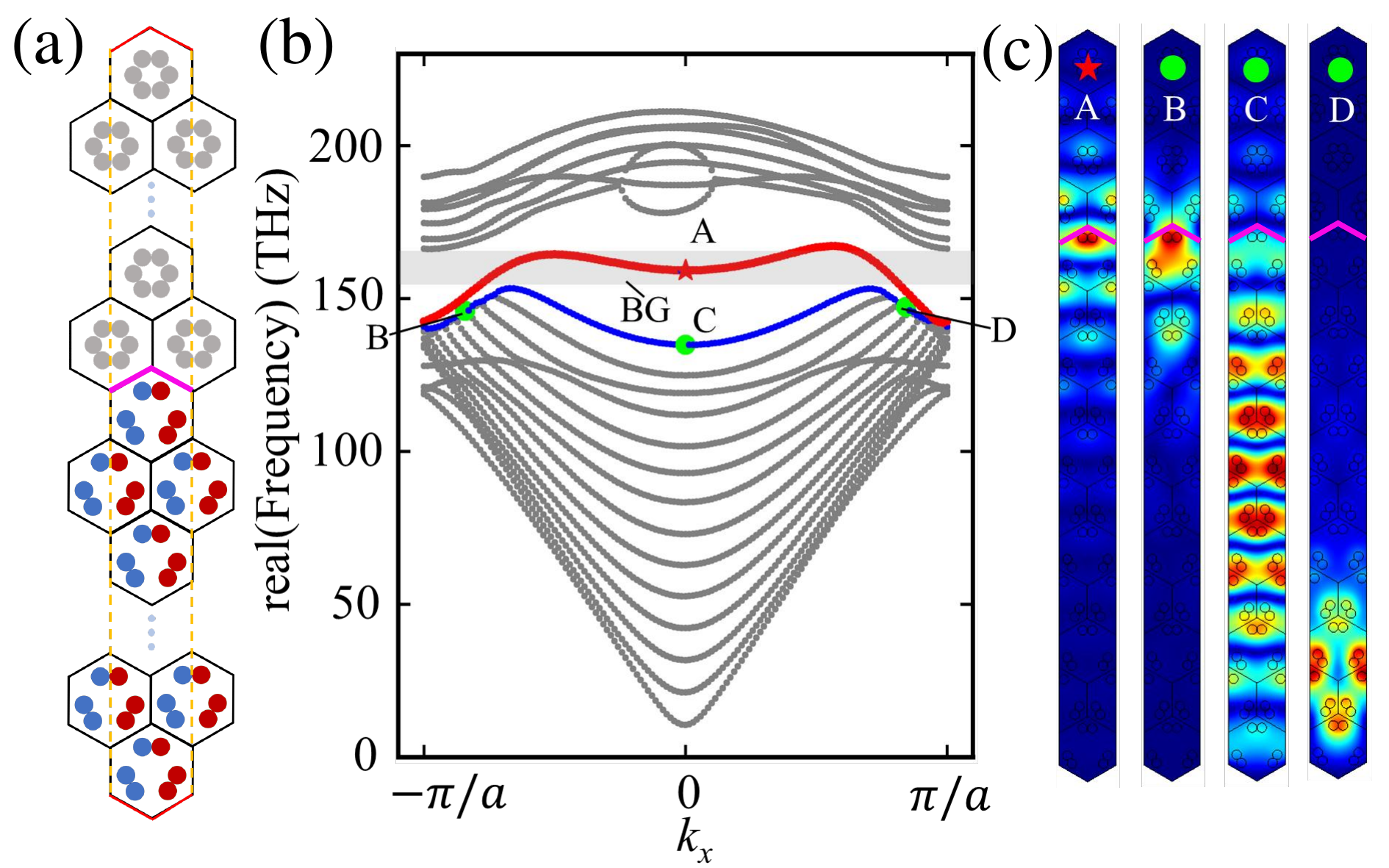}
\caption{\textbf{Topological properties of non-Hermitian photonic crystal supercell with PBC$_x$-OBC$_y$ at $t=1.27$.} (a) Schematic depiction of the ribbon non-Hermitian photonic crystal supercell under PBC$_x$-OBC$_y$. The supercell comprises non-Hermitian extended unit cells and Hermitian shrunken unit cells, with the pink line indicating their domain boundary.
(b) The projected band structure of the supercell. Gray curves represent bulk states (one highlighted in blue), while red curves denote topological edge modes. The gray-shaded region indicates the photonic band gap. The corresponding eigenmode field profiles of the marked states are shown in (c).}\label{fig1Dobc}
\end{figure}
%%%%%%%%%%%%%%%%%%%%%%%%%%%%%%%%%%%%%%%%%%%%

In general, HOT corner modes can arise from breaking protective symmetries, such as time-reversal symmetry, in a first-order topological phase. The resulting symmetry breaking gaps the otherwise gapless edge modes, thereby giving rise to localized higher-order corner/hinge modes within the gap.
In Fig.~\ref{figobc}(b), the topological edge modes can be classified into four complex-conjugate paired parts (E$_{1}$-E$_{4}$) depending on the eigenfrequencies and the eigenmode field profiles. The normalized eigenmode field profiles of the topological edge modes are shown in Figs.~\ref{figedgestates}, illustrating their localization at the four distinct edges of the supercell.
These topological edge modes are localized at the four different edges of the supercell, respectively. 
This is different from the Hermitian case, where the topological edge modes are localized at the boundary with rotation symmetry of the photonic crystal. 
Although topological edge modes are present on all edges, the type-II corner modes does not appear at all corners. In Hermitian systems, type-II corner modes originate from topological edge modes that become spatially confined at corners due to long-range interactions and symmetry constraints~\cite{li2020higher}. In the non-Hermitian system, while the topological edge modes remain distributed along all edges, the NHSE breaks the equivalence between different spatial directions, selecting a preferred corner for localization. As a result, the type-II corner modes localize exclusively at the lower-left corner. Importantly, non-Hermiticity does not change its fundamental origin but instead determines the unique corner through the NHSE. Therefore, the type-II corner mode should be understood as a consequence of the topology of 1D edge modes combined with NHSE.

% By breaking the rotation symmetry of the topological edge modes in our non-Hermitian systems, it is intuitive to reveal the origin of the type-II corner modes. Comparing the eigenmode field profiles of the topological edge modes E$_{1}$, E$_{2}$, and the type-II corner modes CII, we find that long-range interactions induce the localization of the corner modes at the corner where the two edges meet.

%%%%%%%%%%%%%%%%%%%%%%%%%%%%%%%%%%%%
\begin{figure}[b]
\includegraphics[width=0.48\textwidth]{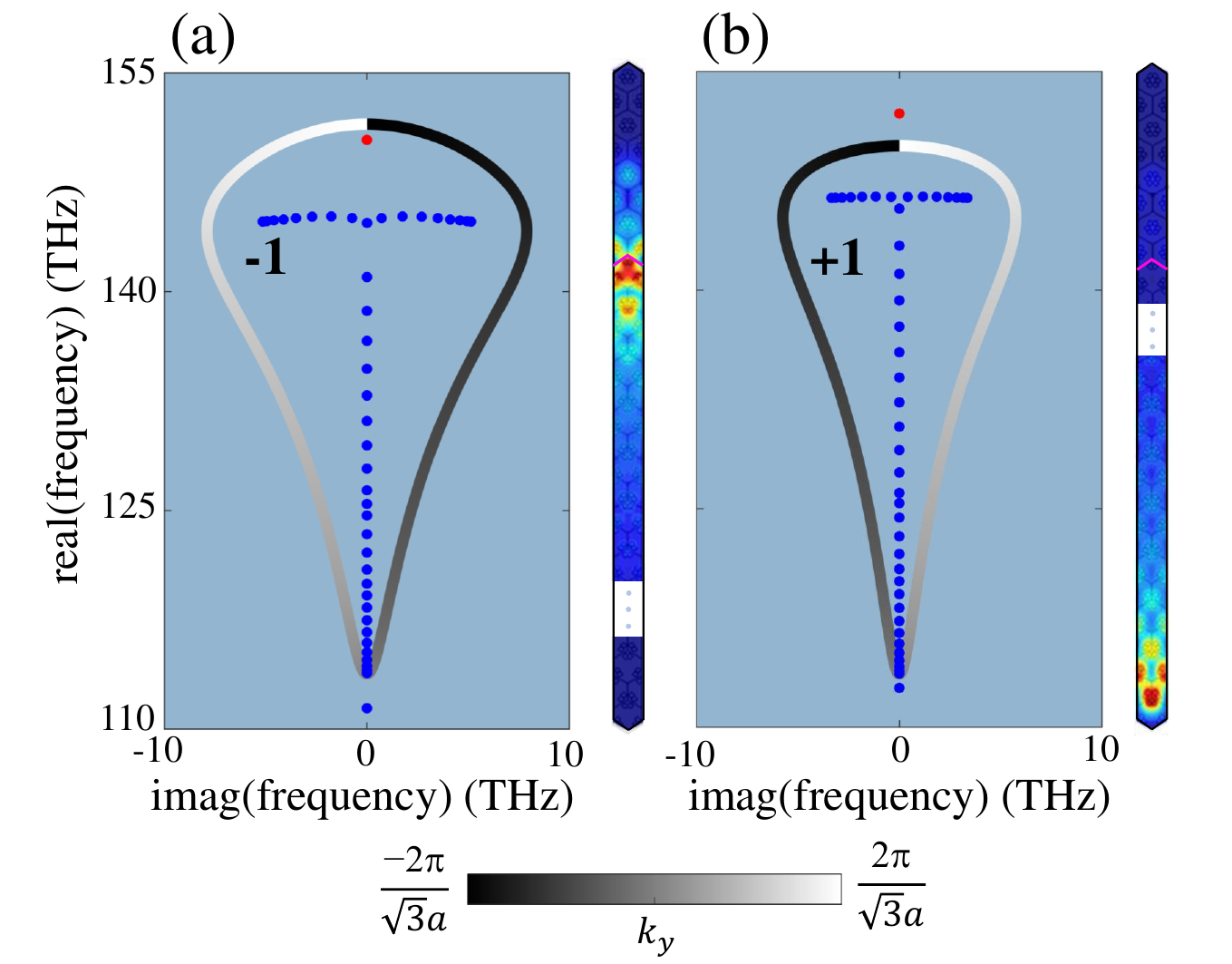}
\caption{\textbf{Point-gap topology of the PBC eigenfrequencies spectra.} (a), (b) The eigenfrequencies spectra at the two $k_x$ points corresponding to green dots B and D in Fig.~\ref{fig1Dobc}(b). The grayscale curves represent the PBC eigenfrequencies spectra, while colored points indicate the OBC$_y$ eigenfrequencies spectra. The corresponding normalized eigenmode field profiles are shown on the right.
}\label{figrealimag}
\end{figure}
%%%%%%%%%%%%%%%%%%%%%%%%%%%%%%%%%%%

To further illustrate the topological properties, we construct a non-Hermitian kagome ribbon supercell with PBC along the $x$ direction and OBC along the $y$ direction (PBC$_x$-OBC$_y$), as shown in Fig.~\ref{fig1Dobc}(a). The supercell consists of non-Hermitian extended unit cells and Hermitian shrunken unit cells. The projected band structure is displayed in Fig.~\ref{fig1Dobc}(b), where a pronounced band gap is highlighted by the gray-shaded region. Within this gap, an isolated band (red) emerges. As shown in the left panel of Fig.~\ref{fig1Dobc}(c), the corresponding eigenmode field profiles are localized at the interface between topologically trivial and nontrivial unit cells, confirming their nature as topological edge modes and further demonstrating the breakdown of the conventional Bloch bulk–boundary correspondence.

In addition to the edge modes, we further examine the eigenmode field profiles of three representative bulk states marked in Fig.~\ref{fig1Dobc}(c). Bulk states with negative momentum $k_x$ are localized along the upper boundary, whereas those with positive $k_x$ localize along the lower boundary. As $k_x$ increases, the localization gradually shifts from the upper to the lower boundary, manifesting a bipolar NHSE along the $y$ direction. An analogous behavior arises along the $x$ direction when imposing PBC$_y$–OBC$_x$, confirming that the non-Hermitian photonic kagome crystal hosts a robust bipolar NHSE in both spatial directions.

The origin and characteristics of the bipolar NHSE can be further elucidated through the point-gap topology of the eigenfrequencies spectrum. Here, we define the topological invariant, namely the winding number $W(k_x, f_{0})$, for a given frequency $f_{0}$ as
\begin{align}
W(k_x,f_{0})&=\oint_{0}^{2\pi}\frac{d k_y}{2\pi i}\frac{\partial}{\partial k_y}\log \det [H(k_x,k_y)-f_{0}].
\tag{6}
\end{align}
The positive (negative) winding numbers correspond to anticlockwise (clockwise) PBC eigenfrequencies variation with momentum. A nonzero winding number of the PBC eigenfrequencies spectrum signals the emergence of the NHSE in the $y$ direction under PBC$_x$-OBC$_y$. Figure~\ref{figrealimag} shows the spectra at $k_x = \pm \tfrac{(2\sqrt{3}-2)\pi}{\sqrt{3}a}$ under PBC (gray curves) and OBC$_y$ (colored points). The two spectra exhibit opposite winding numbers, $W=-1$ and $W=+1$, corresponding to eigenmode localization at opposite boundaries [Figs.~\ref{figrealimag}]. Specifically, a positive (negative) winding number for $k_x>0$ ($k_x<0$) indicates localization at the lower (upper) boundary. This topological correspondence between point-gap winding and boundary localization is not restricted to the $y$ direction: opening boundaries along $x$ likewise produces a bipolar NHSE characterized by the same point-gap topology. Hence, the point-gap topology of the PBC eigenfrequencies spectrum provides the fundamental topological characterization of the bipolar NHSE. 
Non-Hermiticity not only induces intriguing phenomena such as the collapse of PBC eigenfrequencies and the emergence of the bipolar NHSE, but also fundamentally breaks the conventional Bloch bulk-boundary correspondence in photonic crystals.

\section{Conclusion and Discussion}\label{Discussion and Conclusion}

In summary, we have introduced a non-Hermitian photonic kagome crystal that realizes HOT phases in the presence of balanced gain and loss. We showed that non-Hermiticity lifts the degeneracy of corner modes and induces a bipolar NHSE, which drives bulk states to accumulate at opposite corners of the system. This effect provides a direct manifestation of the breakdown of the conventional Bloch bulk-boundary correspondence, as the Bloch band theory no longer captures the eigenfrequencies spectrum and the topology of the non-Hermitian photonic crystal.

By formulating the point-gap topology of the complex eigenfrequencies spectrum, we established the topological origin of the bipolar NHSE and clarified its role in shifting the HOT phase transition. Our findings demonstrate that non-Hermiticity fundamentally alters the spectral topology by collapsing the periodic-boundary spectrum and driving a breakdown of the Bloch bulk-boundary correspondence. The resulting interplay between higher-order topology and the bipolar NHSE establishes a robust mechanism for engineering strongly localized photonic modes, offering new opportunities for controllable light confinement and the design of next-generation photonic devices.

\section*{Acknowledgements}\label{section:Acknowledgements}
We thank Yunjia Zhai and Yiling Zhang for fruitful discussions. This work was supported by Natural Science Foundation of Jiangsu Province (Grant No. BK20231320).

\appendix 
\section*{Appendix}\label{sec:appendix_info}

\begin{figure}[b]
	\includegraphics[width=0.48\textwidth]{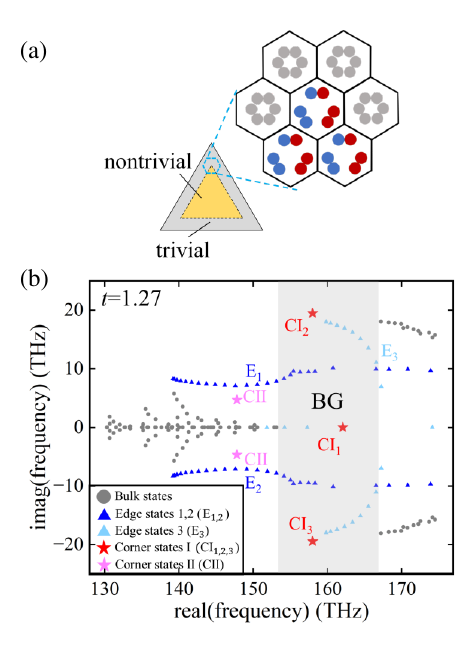}
	\caption
    {\textbf{Complex eigenfrequencies spectrum of triangular supercell ($t = 1.27$) under OBC.} (a) Schematic illustration of the triangular photonic kagome crystal supercell. This supercell consists of 300 non-Hermitian photonic crystal unit cells surrounded by $6$ layers of Hermitian photonic crystal unit cells. (b) The complex eigenfrequencies spectrum for bulk, edge, and corner modes of the triangular supercell under OBC.
    }\label{figA1}
\end{figure}

\begin{figure}[t]
	\includegraphics[width=0.48\textwidth]{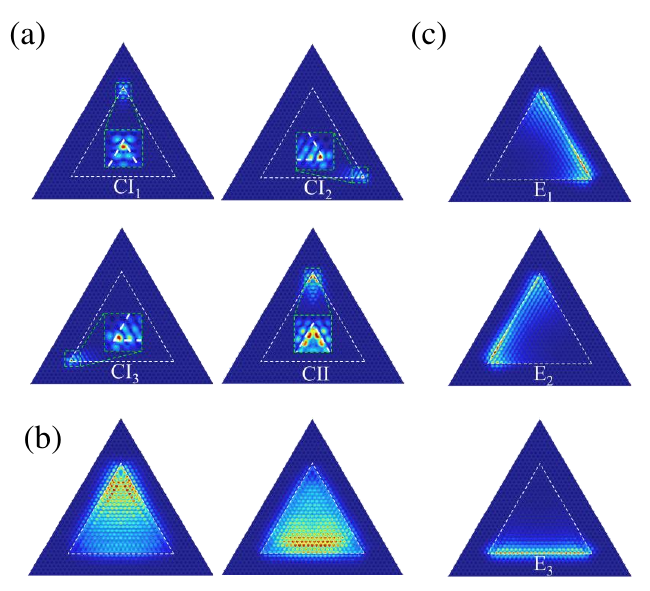}
	\caption
    {\textbf{NHSE, HOT corner modes and edge modes of the non-Hermitian photonic crystal triangular supercell ($t=1.27$) under OBC.} (a) Spatial distribution of the eigenmode field profiles for type-I HOT corner modes CI$_1$, CI$_2$, CI$_3$ and normalized eigenmode field profiles of type-II HOT corner modes CII. (b) Normalized eigenmode field profiles of the bulk states of the supercell. (c) Edge modes E$_1$, E$_2$, and E$_3$ localized at different boundaries of the triangular supercell, respectively.
    }\label{figA2}
\end{figure}

To verify the robustness of our results against variations in geometry, we further investigate a triangular supercell, as shown in Fig.~\ref{figA1}(a). For direct comparison with the rhombic case, we compute the complex eigenfrequencies spectrum of the triangular supercell under the same parameter setting ($t = 1.27$) and the corresponding eigenfrequencies are presented in Fig.~\ref{figA1}(b). The eigenfrequencies spectrum of the triangular supercell also hosts a distinct photonic band gap, further indicates the breakdown of conventional Bloch bulk-boundary correspondence due to the NHSE.

Similarly to the rhombic supercell, isolated eigenfrequencies appear within the band gap, whose eigenmode field profiles confirm their localization at the corners of the triangular supercell, identifying them as HOT corner modes. Specifically, CI$_1$, CI$_2$, and CI$_3$ correspond to type-I corner modes [Fig.~\ref{figA2}(a)]: CI$_1$ exhibits a purely real eigenfrequency and is localized at the top corner, whereas CI$_2$ and CI$_3$ form a pair of complex-conjugate modes localized at the two bottom corners. This demonstrates that non-Hermiticity lifts the degeneracy of type-I corner modes in both eigenfrequencies and spatial localization. In addition, type-II corner modes evolve into paired corner modes with complex-conjugate eigenfrequencies (CII), with eigenmode field profiles localized at the top corner [right panel of Fig.~\ref{figA2}(a)]. As shown in Fig.~\ref{figA2}(b), the bulk states localize at the top corner and the bottom boundary, highlighting the influence of geometry on the bipolar NHSE.

In addition, edge modes are identified by the triangular markers in Fig.~\ref{figA1}(b). Their normalized eigenmode field profiles [Fig.~\ref{figA2}(c)] reveal localization at the three boundaries. Compared with the rhombic case, the eigenfrequencies of edge modes are more sparsely distributed across the complex spectrum. Among them, E$_1$ and E$_2$ form a complex-conjugate pair localized at the left and right edges, while E$_3$ also forms a complex-conjugate pair but localizes along the bottom edge. 

In summary, apart from a more dispersed distribution of edge modes in the complex eigenfrequencies spectrum compared to those in the rhombic supercell, the triangular geometry also exhibits the bipolar NHSE and the breakdown of Bloch bulk-boundary correspondence, confirming that these phenomena are robust against variations in supercell shape.

\bibliography{reference} %title+Hyperlinks

\begin{thebibliography}{99}%
\makeatletter
\providecommand \@ifxundefined [1]{%
 \@ifx{#1\undefined}
}%
\providecommand \@ifnum [1]{%
 \ifnum #1\expandafter \@firstoftwo
 \else \expandafter \@secondoftwo
 \fi
}%
\providecommand \@ifx [1]{%
 \ifx #1\expandafter \@firstoftwo
 \else \expandafter \@secondoftwo
 \fi
}%
\providecommand \natexlab [1]{#1}%
\providecommand \enquote  [1]{``#1''}%
\providecommand \bibnamefont  [1]{#1}%
\providecommand \bibfnamefont [1]{#1}%
\providecommand \citenamefont [1]{#1}%
\providecommand \href@noop [0]{\@secondoftwo}%
\providecommand \href [0]{\begingroup \@sanitize@url \@href}%
\providecommand \@href[1]{\@@startlink{#1}\@@href}%
\providecommand \@@href[1]{\endgroup#1\@@endlink}%
\providecommand \@sanitize@url [0]{\catcode `\\12\catcode `\$12\catcode `\&12\catcode `\#12\catcode `\^12\catcode `\_12\catcode `\%12\relax}%
\providecommand \@@startlink[1]{}%
\providecommand \@@endlink[0]{}%
\providecommand \url  [0]{\begingroup\@sanitize@url \@url }%
\providecommand \@url [1]{\endgroup\@href {#1}{\urlprefix }}%
\providecommand \urlprefix  [0]{URL }%
\providecommand \Eprint [0]{\href }%
\providecommand \doibase [0]{https://doi.org/}%
\providecommand \selectlanguage [0]{\@gobble}%
\providecommand \bibinfo  [0]{\@secondoftwo}%
\providecommand \bibfield  [0]{\@secondoftwo}%
\providecommand \translation [1]{[#1]}%
\providecommand \BibitemOpen [0]{}%
\providecommand \bibitemStop [0]{}%
\providecommand \bibitemNoStop [0]{.\EOS\space}%
\providecommand \EOS [0]{\spacefactor3000\relax}%
\providecommand \BibitemShut  [1]{\csname bibitem#1\endcsname}%
\let\auto@bib@innerbib\@empty
%</preamble>
\bibitem [{\citenamefont {Bender}\ and\ \citenamefont {Boettcher}(1998)}]{PhysRevLett.80.5243}%
  \BibitemOpen
  \bibfield  {author} {\bibinfo {author} {\bibfnamefont {C.~M.}\ \bibnamefont {Bender}}\ and\ \bibinfo {author} {\bibfnamefont {S.}~\bibnamefont {Boettcher}},\ }\bibfield  {title} {\bibinfo {title} {{Real Spectra in Non-Hermitian Hamiltonians Having $\mathcal{P}\mathcal{T}$ Symmetry}},\ }\href {https://doi.org/10.1103/PhysRevLett.80.5243} {\bibfield  {journal} {\bibinfo  {journal} {Phys. Rev. Lett.}\ }\textbf {\bibinfo {volume} {80}},\ \bibinfo {pages} {5243} (\bibinfo {year} {1998})}\BibitemShut {NoStop}%
\bibitem [{\citenamefont {Zeuner}\ \emph {et~al.}(2015)\citenamefont {Zeuner}, \citenamefont {Rechtsman}, \citenamefont {Plotnik}, \citenamefont {Lumer}, \citenamefont {Nolte}, \citenamefont {Rudner}, \citenamefont {Segev},\ and\ \citenamefont {Szameit}}]{PhysRevLett.115.040402}%
  \BibitemOpen
  \bibfield  {author} {\bibinfo {author} {\bibfnamefont {J.~M.}\ \bibnamefont {Zeuner}}, \bibinfo {author} {\bibfnamefont {M.~C.}\ \bibnamefont {Rechtsman}}, \bibinfo {author} {\bibfnamefont {Y.}~\bibnamefont {Plotnik}}, \bibinfo {author} {\bibfnamefont {Y.}~\bibnamefont {Lumer}}, \bibinfo {author} {\bibfnamefont {S.}~\bibnamefont {Nolte}}, \bibinfo {author} {\bibfnamefont {M.~S.}\ \bibnamefont {Rudner}}, \bibinfo {author} {\bibfnamefont {M.}~\bibnamefont {Segev}},\ and\ \bibinfo {author} {\bibfnamefont {A.}~\bibnamefont {Szameit}},\ }\bibfield  {title} {\bibinfo {title} {{Observation of a Topological Transition in the Bulk of a Non-Hermitian System}},\ }\href {https://doi.org/10.1103/PhysRevLett.115.040402} {\bibfield  {journal} {\bibinfo  {journal} {Phys. Rev. Lett.}\ }\textbf {\bibinfo {volume} {115}},\ \bibinfo {pages} {040402} (\bibinfo {year} {2015})}\BibitemShut {NoStop}%
\bibitem [{\citenamefont {Leykam}\ \emph {et~al.}(2017)\citenamefont {Leykam}, \citenamefont {Bliokh}, \citenamefont {Huang}, \citenamefont {Chong},\ and\ \citenamefont {Nori}}]{PhysRevLett.118.040401}%
  \BibitemOpen
  \bibfield  {author} {\bibinfo {author} {\bibfnamefont {D.}~\bibnamefont {Leykam}}, \bibinfo {author} {\bibfnamefont {K.~Y.}\ \bibnamefont {Bliokh}}, \bibinfo {author} {\bibfnamefont {C.}~\bibnamefont {Huang}}, \bibinfo {author} {\bibfnamefont {Y.~D.}\ \bibnamefont {Chong}},\ and\ \bibinfo {author} {\bibfnamefont {F.}~\bibnamefont {Nori}},\ }\bibfield  {title} {\bibinfo {title} {{Edge Modes, Degeneracies, and Topological Numbers in Non-Hermitian Systems}},\ }\href {https://doi.org/10.1103/PhysRevLett.118.040401} {\bibfield  {journal} {\bibinfo  {journal} {Phys. Rev. Lett.}\ }\textbf {\bibinfo {volume} {118}},\ \bibinfo {pages} {040401} (\bibinfo {year} {2017})}\BibitemShut {NoStop}%
\bibitem [{\citenamefont {Lee}\ \emph {et~al.}(2019)\citenamefont {Lee}, \citenamefont {Ahn}, \citenamefont {Zhou},\ and\ \citenamefont {Vishwanath}}]{PhysRevLett.123.206404}%
  \BibitemOpen
  \bibfield  {author} {\bibinfo {author} {\bibfnamefont {J.~Y.}\ \bibnamefont {Lee}}, \bibinfo {author} {\bibfnamefont {J.}~\bibnamefont {Ahn}}, \bibinfo {author} {\bibfnamefont {H.}~\bibnamefont {Zhou}},\ and\ \bibinfo {author} {\bibfnamefont {A.}~\bibnamefont {Vishwanath}},\ }\bibfield  {title} {\bibinfo {title} {{Topological Correspondence between Hermitian and Non-Hermitian Systems: Anomalous Dynamics}},\ }\href {https://doi.org/10.1103/PhysRevLett.123.206404} {\bibfield  {journal} {\bibinfo  {journal} {Phys. Rev. Lett.}\ }\textbf {\bibinfo {volume} {123}},\ \bibinfo {pages} {206404} (\bibinfo {year} {2019})}\BibitemShut {NoStop}%
\bibitem [{\citenamefont {Song}\ \emph {et~al.}(2019{\natexlab{a}})\citenamefont {Song}, \citenamefont {Yao},\ and\ \citenamefont {Wang}}]{PhysRevLett.123.246801}%
  \BibitemOpen
  \bibfield  {author} {\bibinfo {author} {\bibfnamefont {F.}~\bibnamefont {Song}}, \bibinfo {author} {\bibfnamefont {S.}~\bibnamefont {Yao}},\ and\ \bibinfo {author} {\bibfnamefont {Z.}~\bibnamefont {Wang}},\ }\bibfield  {title} {\bibinfo {title} {{Non-Hermitian Topological Invariants in Real Space}},\ }\href {https://doi.org/10.1103/PhysRevLett.123.246801} {\bibfield  {journal} {\bibinfo  {journal} {Phys. Rev. Lett.}\ }\textbf {\bibinfo {volume} {123}},\ \bibinfo {pages} {246801} (\bibinfo {year} {2019}{\natexlab{a}})}\BibitemShut {NoStop}%
\bibitem [{\citenamefont {Lee}\ and\ \citenamefont {Thomale}(2019)}]{PhysRevB.99.201103}%
  \BibitemOpen
  \bibfield  {author} {\bibinfo {author} {\bibfnamefont {C.~H.}\ \bibnamefont {Lee}}\ and\ \bibinfo {author} {\bibfnamefont {R.}~\bibnamefont {Thomale}},\ }\bibfield  {title} {\bibinfo {title} {{Anatomy of skin modes and topology in non-Hermitian systems}},\ }\href {https://doi.org/10.1103/PhysRevB.99.201103} {\bibfield  {journal} {\bibinfo  {journal} {Phys. Rev. B}\ }\textbf {\bibinfo {volume} {99}},\ \bibinfo {pages} {201103} (\bibinfo {year} {2019})}\BibitemShut {NoStop}%
\bibitem [{\citenamefont {Ashida}\ \emph {et~al.}(2020)\citenamefont {Ashida}, \citenamefont {Gong},\ and\ \citenamefont {Ueda}}]{Ashida02072020}%
  \BibitemOpen
  \bibfield  {author} {\bibinfo {author} {\bibfnamefont {Y.}~\bibnamefont {Ashida}}, \bibinfo {author} {\bibfnamefont {Z.}~\bibnamefont {Gong}},\ and\ \bibinfo {author} {\bibfnamefont {M.}~\bibnamefont {Ueda}},\ }\bibfield  {title} {\bibinfo {title} {Non-hermitian physics},\ }\href {https://doi.org/10.1080/00018732.2021.1876991} {\bibfield  {journal} {\bibinfo  {journal} {Adv. Phys.}\ }\textbf {\bibinfo {volume} {69}},\ \bibinfo {pages} {249} (\bibinfo {year} {2020})}\BibitemShut {NoStop}%
\bibitem [{\citenamefont {Bergholtz}\ \emph {et~al.}(2021)\citenamefont {Bergholtz}, \citenamefont {Budich},\ and\ \citenamefont {Kunst}}]{RevModPhys.93.015005}%
  \BibitemOpen
  \bibfield  {author} {\bibinfo {author} {\bibfnamefont {E.~J.}\ \bibnamefont {Bergholtz}}, \bibinfo {author} {\bibfnamefont {J.~C.}\ \bibnamefont {Budich}},\ and\ \bibinfo {author} {\bibfnamefont {F.~K.}\ \bibnamefont {Kunst}},\ }\bibfield  {title} {\bibinfo {title} {{Exceptional topology of non-Hermitian systems}},\ }\href {https://doi.org/10.1103/RevModPhys.93.015005} {\bibfield  {journal} {\bibinfo  {journal} {Rev. Mod. Phys.}\ }\textbf {\bibinfo {volume} {93}},\ \bibinfo {pages} {015005} (\bibinfo {year} {2021})}\BibitemShut {NoStop}%
\bibitem [{\citenamefont {Kawabata}\ \emph {et~al.}(2022)\citenamefont {Kawabata}, \citenamefont {Shiozaki},\ and\ \citenamefont {Ryu}}]{PhysRevB.105.165137}%
  \BibitemOpen
  \bibfield  {author} {\bibinfo {author} {\bibfnamefont {K.}~\bibnamefont {Kawabata}}, \bibinfo {author} {\bibfnamefont {K.}~\bibnamefont {Shiozaki}},\ and\ \bibinfo {author} {\bibfnamefont {S.}~\bibnamefont {Ryu}},\ }\bibfield  {title} {\bibinfo {title} {{Many-body topology of non-Hermitian systems}},\ }\href {https://doi.org/10.1103/PhysRevB.105.165137} {\bibfield  {journal} {\bibinfo  {journal} {Phys. Rev. B}\ }\textbf {\bibinfo {volume} {105}},\ \bibinfo {pages} {165137} (\bibinfo {year} {2022})}\BibitemShut {NoStop}%
\bibitem [{\citenamefont {Edvardsson}\ and\ \citenamefont {Ardonne}(2022)}]{PhysRevB.106.115107}%
  \BibitemOpen
  \bibfield  {author} {\bibinfo {author} {\bibfnamefont {E.}~\bibnamefont {Edvardsson}}\ and\ \bibinfo {author} {\bibfnamefont {E.}~\bibnamefont {Ardonne}},\ }\bibfield  {title} {\bibinfo {title} {{Sensitivity of non-Hermitian systems}},\ }\href {https://doi.org/10.1103/PhysRevB.106.115107} {\bibfield  {journal} {\bibinfo  {journal} {Phys. Rev. B}\ }\textbf {\bibinfo {volume} {106}},\ \bibinfo {pages} {115107} (\bibinfo {year} {2022})}\BibitemShut {NoStop}%
\bibitem [{\citenamefont {Shi}\ \emph {et~al.}(2024)\citenamefont {Shi}, \citenamefont {Peng}, \citenamefont {Peng}, \citenamefont {Chen},\ and\ \citenamefont {Liu}}]{PhysRevB.110.014106}%
  \BibitemOpen
  \bibfield  {author} {\bibinfo {author} {\bibfnamefont {A.}~\bibnamefont {Shi}}, \bibinfo {author} {\bibfnamefont {Y.}~\bibnamefont {Peng}}, \bibinfo {author} {\bibfnamefont {P.}~\bibnamefont {Peng}}, \bibinfo {author} {\bibfnamefont {J.}~\bibnamefont {Chen}},\ and\ \bibinfo {author} {\bibfnamefont {J.}~\bibnamefont {Liu}},\ }\bibfield  {title} {\bibinfo {title} {Delocalization of higher-order topological states in higher-dimensional non-hermitian quasicrystals},\ }\href {https://doi.org/10.1103/PhysRevB.110.014106} {\bibfield  {journal} {\bibinfo  {journal} {Phys. Rev. B}\ }\textbf {\bibinfo {volume} {110}},\ \bibinfo {pages} {014106} (\bibinfo {year} {2024})}\BibitemShut {NoStop}%
\bibitem [{\citenamefont {Rangi}\ \emph {et~al.}(2024)\citenamefont {Rangi}, \citenamefont {Tam},\ and\ \citenamefont {Moreno}}]{PhysRevB.109.064203}%
  \BibitemOpen
  \bibfield  {author} {\bibinfo {author} {\bibfnamefont {C.}~\bibnamefont {Rangi}}, \bibinfo {author} {\bibfnamefont {K.-M.}\ \bibnamefont {Tam}},\ and\ \bibinfo {author} {\bibfnamefont {J.}~\bibnamefont {Moreno}},\ }\bibfield  {title} {\bibinfo {title} {Engineering a non-hermitian second-order topological insulator state in quasicrystals},\ }\href {https://doi.org/10.1103/PhysRevB.109.064203} {\bibfield  {journal} {\bibinfo  {journal} {Phys. Rev. B}\ }\textbf {\bibinfo {volume} {109}},\ \bibinfo {pages} {064203} (\bibinfo {year} {2024})}\BibitemShut {NoStop}%
\bibitem [{\citenamefont {Ji}\ \emph {et~al.}(2024)\citenamefont {Ji}, \citenamefont {Ding}, \citenamefont {Chen},\ and\ \citenamefont {Yang}}]{PhysRevB.109.125420}%
  \BibitemOpen
  \bibfield  {author} {\bibinfo {author} {\bibfnamefont {X.}~\bibnamefont {Ji}}, \bibinfo {author} {\bibfnamefont {W.}~\bibnamefont {Ding}}, \bibinfo {author} {\bibfnamefont {Y.}~\bibnamefont {Chen}},\ and\ \bibinfo {author} {\bibfnamefont {X.}~\bibnamefont {Yang}},\ }\bibfield  {title} {\bibinfo {title} {Non-hermitian second-order topological superconductors},\ }\href {https://doi.org/10.1103/PhysRevB.109.125420} {\bibfield  {journal} {\bibinfo  {journal} {Phys. Rev. B}\ }\textbf {\bibinfo {volume} {109}},\ \bibinfo {pages} {125420} (\bibinfo {year} {2024})}\BibitemShut {NoStop}%
\bibitem [{\citenamefont {Li}\ \emph {et~al.}(2025{\natexlab{a}})\citenamefont {Li}, \citenamefont {Wei}, \citenamefont {Wu}, \citenamefont {Ruan}, \citenamefont {Chen}, \citenamefont {Lee},\ and\ \citenamefont {Ni}}]{PhysRevB.111.075132}%
  \BibitemOpen
  \bibfield  {author} {\bibinfo {author} {\bibfnamefont {L.}~\bibnamefont {Li}}, \bibinfo {author} {\bibfnamefont {Y.}~\bibnamefont {Wei}}, \bibinfo {author} {\bibfnamefont {G.}~\bibnamefont {Wu}}, \bibinfo {author} {\bibfnamefont {Y.}~\bibnamefont {Ruan}}, \bibinfo {author} {\bibfnamefont {S.}~\bibnamefont {Chen}}, \bibinfo {author} {\bibfnamefont {C.~H.}\ \bibnamefont {Lee}},\ and\ \bibinfo {author} {\bibfnamefont {Z.}~\bibnamefont {Ni}},\ }\bibfield  {title} {\bibinfo {title} {Exact solutions disentangle higher-order topology in two-dimensional non-hermitian lattices},\ }\href {https://doi.org/10.1103/PhysRevB.111.075132} {\bibfield  {journal} {\bibinfo  {journal} {Phys. Rev. B}\ }\textbf {\bibinfo {volume} {111}},\ \bibinfo {pages} {075132} (\bibinfo {year} {2025}{\natexlab{a}})}\BibitemShut {NoStop}%
\bibitem [{\citenamefont {Shi}\ \emph {et~al.}(2025)\citenamefont {Shi}, \citenamefont {Bao}, \citenamefont {Peng}, \citenamefont {Ning}, \citenamefont {Wang},\ and\ \citenamefont {Liu}}]{PhysRevB.111.094109}%
  \BibitemOpen
  \bibfield  {author} {\bibinfo {author} {\bibfnamefont {A.}~\bibnamefont {Shi}}, \bibinfo {author} {\bibfnamefont {L.}~\bibnamefont {Bao}}, \bibinfo {author} {\bibfnamefont {P.}~\bibnamefont {Peng}}, \bibinfo {author} {\bibfnamefont {J.}~\bibnamefont {Ning}}, \bibinfo {author} {\bibfnamefont {Z.}~\bibnamefont {Wang}},\ and\ \bibinfo {author} {\bibfnamefont {J.}~\bibnamefont {Liu}},\ }\bibfield  {title} {\bibinfo {title} {Non-hermitian floquet higher-order topological states in two-dimensional quasicrystals},\ }\href {https://doi.org/10.1103/PhysRevB.111.094109} {\bibfield  {journal} {\bibinfo  {journal} {Phys. Rev. B}\ }\textbf {\bibinfo {volume} {111}},\ \bibinfo {pages} {094109} (\bibinfo {year} {2025})}\BibitemShut {NoStop}%
\bibitem [{\citenamefont {Ji}\ \emph {et~al.}(2025)\citenamefont {Ji}, \citenamefont {Geng}, \citenamefont {Akhtar},\ and\ \citenamefont {Yang}}]{PhysRevB.111.195419}%
  \BibitemOpen
  \bibfield  {author} {\bibinfo {author} {\bibfnamefont {X.}~\bibnamefont {Ji}}, \bibinfo {author} {\bibfnamefont {H.}~\bibnamefont {Geng}}, \bibinfo {author} {\bibfnamefont {N.}~\bibnamefont {Akhtar}},\ and\ \bibinfo {author} {\bibfnamefont {X.}~\bibnamefont {Yang}},\ }\bibfield  {title} {\bibinfo {title} {{Floquet engineering of point-gapped topological superconductors}},\ }\href {https://doi.org/10.1103/PhysRevB.111.195419} {\bibfield  {journal} {\bibinfo  {journal} {Phys. Rev. B}\ }\textbf {\bibinfo {volume} {111}},\ \bibinfo {pages} {195419} (\bibinfo {year} {2025})}\BibitemShut {NoStop}%
\bibitem [{\citenamefont {Yao}\ and\ \citenamefont {Wang}(2018)}]{yao2018edge}%
  \BibitemOpen
  \bibfield  {author} {\bibinfo {author} {\bibfnamefont {S.}~\bibnamefont {Yao}}\ and\ \bibinfo {author} {\bibfnamefont {Z.}~\bibnamefont {Wang}},\ }\bibfield  {title} {\bibinfo {title} {{Edge states and topological invariants of non-Hermitian systems}},\ }\href {https://doi.org/10.1103/PhysRevLett.121.086803} {\bibfield  {journal} {\bibinfo  {journal} {Phys. Rev. Lett.}\ }\textbf {\bibinfo {volume} {121}},\ \bibinfo {pages} {086803} (\bibinfo {year} {2018})}\BibitemShut {NoStop}%
\bibitem [{\citenamefont {Song}\ \emph {et~al.}(2019{\natexlab{b}})\citenamefont {Song}, \citenamefont {Yao},\ and\ \citenamefont {Wang}}]{PhysRevLett.123.170401}%
  \BibitemOpen
  \bibfield  {author} {\bibinfo {author} {\bibfnamefont {F.}~\bibnamefont {Song}}, \bibinfo {author} {\bibfnamefont {S.}~\bibnamefont {Yao}},\ and\ \bibinfo {author} {\bibfnamefont {Z.}~\bibnamefont {Wang}},\ }\bibfield  {title} {\bibinfo {title} {{Non-Hermitian Skin Effect and Chiral Damping in Open Quantum Systems}},\ }\href {https://doi.org/10.1103/PhysRevLett.123.170401} {\bibfield  {journal} {\bibinfo  {journal} {Phys. Rev. Lett.}\ }\textbf {\bibinfo {volume} {123}},\ \bibinfo {pages} {170401} (\bibinfo {year} {2019}{\natexlab{b}})}\BibitemShut {NoStop}%
\bibitem [{\citenamefont {Longhi}(2019{\natexlab{a}})}]{PhysRevResearch.1.023013}%
  \BibitemOpen
  \bibfield  {author} {\bibinfo {author} {\bibfnamefont {S.}~\bibnamefont {Longhi}},\ }\bibfield  {title} {\bibinfo {title} {{Probing non-Hermitian skin effect and non-Bloch phase transitions}},\ }\href {https://doi.org/10.1103/PhysRevResearch.1.023013} {\bibfield  {journal} {\bibinfo  {journal} {Phys. Rev. Res.}\ }\textbf {\bibinfo {volume} {1}},\ \bibinfo {pages} {023013} (\bibinfo {year} {2019}{\natexlab{a}})}\BibitemShut {NoStop}%
\bibitem [{\citenamefont {Zhang}\ \emph {et~al.}(2020)\citenamefont {Zhang}, \citenamefont {Yang},\ and\ \citenamefont {Fang}}]{PhysRevLett.125.126402}%
  \BibitemOpen
  \bibfield  {author} {\bibinfo {author} {\bibfnamefont {K.}~\bibnamefont {Zhang}}, \bibinfo {author} {\bibfnamefont {Z.}~\bibnamefont {Yang}},\ and\ \bibinfo {author} {\bibfnamefont {C.}~\bibnamefont {Fang}},\ }\bibfield  {title} {\bibinfo {title} {{Correspondence between Winding Numbers and Skin Modes in Non-Hermitian Systems}},\ }\href {https://doi.org/10.1103/PhysRevLett.125.126402} {\bibfield  {journal} {\bibinfo  {journal} {Phys. Rev. Lett.}\ }\textbf {\bibinfo {volume} {125}},\ \bibinfo {pages} {126402} (\bibinfo {year} {2020})}\BibitemShut {NoStop}%
\bibitem [{\citenamefont {Li}\ \emph {et~al.}(2020{\natexlab{a}})\citenamefont {Li}, \citenamefont {Lee}, \citenamefont {Mu},\ and\ \citenamefont {Gong}}]{li2020critical}%
  \BibitemOpen
  \bibfield  {author} {\bibinfo {author} {\bibfnamefont {L.}~\bibnamefont {Li}}, \bibinfo {author} {\bibfnamefont {C.~H.}\ \bibnamefont {Lee}}, \bibinfo {author} {\bibfnamefont {S.}~\bibnamefont {Mu}},\ and\ \bibinfo {author} {\bibfnamefont {J.}~\bibnamefont {Gong}},\ }\bibfield  {title} {\bibinfo {title} {{Critical non-Hermitian skin effect}},\ }\href {https://doi.org/10.1038/s41467-020-18917-4} {\bibfield  {journal} {\bibinfo  {journal} {Nat. Commun.}\ }\textbf {\bibinfo {volume} {11}},\ \bibinfo {pages} {5491} (\bibinfo {year} {2020}{\natexlab{a}})}\BibitemShut {NoStop}%
\bibitem [{\citenamefont {Okuma}\ \emph {et~al.}(2020)\citenamefont {Okuma}, \citenamefont {Kawabata}, \citenamefont {Shiozaki},\ and\ \citenamefont {Sato}}]{PhysRevLett.124.086801}%
  \BibitemOpen
  \bibfield  {author} {\bibinfo {author} {\bibfnamefont {N.}~\bibnamefont {Okuma}}, \bibinfo {author} {\bibfnamefont {K.}~\bibnamefont {Kawabata}}, \bibinfo {author} {\bibfnamefont {K.}~\bibnamefont {Shiozaki}},\ and\ \bibinfo {author} {\bibfnamefont {M.}~\bibnamefont {Sato}},\ }\bibfield  {title} {\bibinfo {title} {{Topological Origin of Non-Hermitian Skin Effects}},\ }\href {https://doi.org/10.1103/PhysRevLett.124.086801} {\bibfield  {journal} {\bibinfo  {journal} {Phys. Rev. Lett.}\ }\textbf {\bibinfo {volume} {124}},\ \bibinfo {pages} {086801} (\bibinfo {year} {2020})}\BibitemShut {NoStop}%
\bibitem [{\citenamefont {Scheibner}\ \emph {et~al.}(2020)\citenamefont {Scheibner}, \citenamefont {Irvine},\ and\ \citenamefont {Vitelli}}]{PhysRevLett.125.118001}%
  \BibitemOpen
  \bibfield  {author} {\bibinfo {author} {\bibfnamefont {C.}~\bibnamefont {Scheibner}}, \bibinfo {author} {\bibfnamefont {W.~T.~M.}\ \bibnamefont {Irvine}},\ and\ \bibinfo {author} {\bibfnamefont {V.}~\bibnamefont {Vitelli}},\ }\bibfield  {title} {\bibinfo {title} {{Non-Hermitian Band Topology and Skin Modes in Active Elastic Media}},\ }\href {https://doi.org/10.1103/PhysRevLett.125.118001} {\bibfield  {journal} {\bibinfo  {journal} {Phys. Rev. Lett.}\ }\textbf {\bibinfo {volume} {125}},\ \bibinfo {pages} {118001} (\bibinfo {year} {2020})}\BibitemShut {NoStop}%
\bibitem [{\citenamefont {Kawabata}\ \emph {et~al.}(2020)\citenamefont {Kawabata}, \citenamefont {Sato},\ and\ \citenamefont {Shiozaki}}]{PhysRevB.102.205118}%
  \BibitemOpen
  \bibfield  {author} {\bibinfo {author} {\bibfnamefont {K.}~\bibnamefont {Kawabata}}, \bibinfo {author} {\bibfnamefont {M.}~\bibnamefont {Sato}},\ and\ \bibinfo {author} {\bibfnamefont {K.}~\bibnamefont {Shiozaki}},\ }\bibfield  {title} {\bibinfo {title} {{Higher-order non-Hermitian skin effect}},\ }\href {https://doi.org/10.1103/PhysRevB.102.205118} {\bibfield  {journal} {\bibinfo  {journal} {Phys. Rev. B}\ }\textbf {\bibinfo {volume} {102}},\ \bibinfo {pages} {205118} (\bibinfo {year} {2020})}\BibitemShut {NoStop}%
\bibitem [{\citenamefont {Zhang}\ \emph {et~al.}(2021)\citenamefont {Zhang}, \citenamefont {Tian}, \citenamefont {Jiang}, \citenamefont {Lu},\ and\ \citenamefont {Chen}}]{zhang2021observation}%
  \BibitemOpen
  \bibfield  {author} {\bibinfo {author} {\bibfnamefont {X.}~\bibnamefont {Zhang}}, \bibinfo {author} {\bibfnamefont {Y.}~\bibnamefont {Tian}}, \bibinfo {author} {\bibfnamefont {J.-H.}\ \bibnamefont {Jiang}}, \bibinfo {author} {\bibfnamefont {M.-H.}\ \bibnamefont {Lu}},\ and\ \bibinfo {author} {\bibfnamefont {Y.-F.}\ \bibnamefont {Chen}},\ }\bibfield  {title} {\bibinfo {title} {{Observation of higher-order non-Hermitian skin effect}},\ }\href {https://doi.org/10.1038/s41467-021-25716-y} {\bibfield  {journal} {\bibinfo  {journal} {Nat. Commun.}\ }\textbf {\bibinfo {volume} {12}},\ \bibinfo {pages} {5377} (\bibinfo {year} {2021})}\BibitemShut {NoStop}%
\bibitem [{\citenamefont {Zhang}\ \emph {et~al.}(2022)\citenamefont {Zhang}, \citenamefont {Yang},\ and\ \citenamefont {Fang}}]{zhang2022universal}%
  \BibitemOpen
  \bibfield  {author} {\bibinfo {author} {\bibfnamefont {K.}~\bibnamefont {Zhang}}, \bibinfo {author} {\bibfnamefont {Z.}~\bibnamefont {Yang}},\ and\ \bibinfo {author} {\bibfnamefont {C.}~\bibnamefont {Fang}},\ }\bibfield  {title} {\bibinfo {title} {{Universal non-Hermitian skin effect in two and higher dimensions}},\ }\href {https://doi.org/10.1038/s41467-022-30161-6} {\bibfield  {journal} {\bibinfo  {journal} {Nat. Commun.}\ }\textbf {\bibinfo {volume} {13}},\ \bibinfo {pages} {2496} (\bibinfo {year} {2022})}\BibitemShut {NoStop}%
\bibitem [{\citenamefont {Li}\ \emph {et~al.}(2022)\citenamefont {Li}, \citenamefont {Ji}, \citenamefont {Chen}, \citenamefont {Yan},\ and\ \citenamefont {Yang}}]{PhysRevB.106.195425}%
  \BibitemOpen
  \bibfield  {author} {\bibinfo {author} {\bibfnamefont {Y.}~\bibnamefont {Li}}, \bibinfo {author} {\bibfnamefont {X.}~\bibnamefont {Ji}}, \bibinfo {author} {\bibfnamefont {Y.}~\bibnamefont {Chen}}, \bibinfo {author} {\bibfnamefont {X.}~\bibnamefont {Yan}},\ and\ \bibinfo {author} {\bibfnamefont {X.}~\bibnamefont {Yang}},\ }\bibfield  {title} {\bibinfo {title} {{Topological energy braiding of non-Bloch bands}},\ }\href {https://doi.org/10.1103/PhysRevB.106.195425} {\bibfield  {journal} {\bibinfo  {journal} {Phys. Rev. B}\ }\textbf {\bibinfo {volume} {106}},\ \bibinfo {pages} {195425} (\bibinfo {year} {2022})}\BibitemShut {NoStop}%
\bibitem [{\citenamefont {Hu}(2025)}]{HU202551}%
  \BibitemOpen
  \bibfield  {author} {\bibinfo {author} {\bibfnamefont {H.}~\bibnamefont {Hu}},\ }\bibfield  {title} {\bibinfo {title} {{Topological origin of non-Hermitian skin effect in higher dimensions and uniform spectra}},\ }\href {https://doi.org/https://doi.org/10.1016/j.scib.2024.07.022} {\bibfield  {journal} {\bibinfo  {journal} {Sci. Bull.}\ }\textbf {\bibinfo {volume} {70}},\ \bibinfo {pages} {51} (\bibinfo {year} {2025})}\BibitemShut {NoStop}%
\bibitem [{\citenamefont {Okugawa}\ \emph {et~al.}(2020)\citenamefont {Okugawa}, \citenamefont {Takahashi},\ and\ \citenamefont {Yokomizo}}]{PhysRevB.102.241202}%
  \BibitemOpen
  \bibfield  {author} {\bibinfo {author} {\bibfnamefont {R.}~\bibnamefont {Okugawa}}, \bibinfo {author} {\bibfnamefont {R.}~\bibnamefont {Takahashi}},\ and\ \bibinfo {author} {\bibfnamefont {K.}~\bibnamefont {Yokomizo}},\ }\bibfield  {title} {\bibinfo {title} {Second-order topological non-hermitian skin effects},\ }\href {https://doi.org/10.1103/PhysRevB.102.241202} {\bibfield  {journal} {\bibinfo  {journal} {Phys. Rev. B}\ }\textbf {\bibinfo {volume} {102}},\ \bibinfo {pages} {241202} (\bibinfo {year} {2020})}\BibitemShut {NoStop}%
\bibitem [{\citenamefont {Yokomizo}\ and\ \citenamefont {Murakami}(2021)}]{PhysRevB.104.165117}%
  \BibitemOpen
  \bibfield  {author} {\bibinfo {author} {\bibfnamefont {K.}~\bibnamefont {Yokomizo}}\ and\ \bibinfo {author} {\bibfnamefont {S.}~\bibnamefont {Murakami}},\ }\bibfield  {title} {\bibinfo {title} {Scaling rule for the critical non-hermitian skin effect},\ }\href {https://doi.org/10.1103/PhysRevB.104.165117} {\bibfield  {journal} {\bibinfo  {journal} {Phys. Rev. B}\ }\textbf {\bibinfo {volume} {104}},\ \bibinfo {pages} {165117} (\bibinfo {year} {2021})}\BibitemShut {NoStop}%
\bibitem [{\citenamefont {Gu}\ \emph {et~al.}(2022)\citenamefont {Gu}, \citenamefont {Gao}, \citenamefont {Xue}, \citenamefont {Li}, \citenamefont {Su},\ and\ \citenamefont {Zhu}}]{gu2022transient}%
  \BibitemOpen
  \bibfield  {author} {\bibinfo {author} {\bibfnamefont {Z.}~\bibnamefont {Gu}}, \bibinfo {author} {\bibfnamefont {H.}~\bibnamefont {Gao}}, \bibinfo {author} {\bibfnamefont {H.}~\bibnamefont {Xue}}, \bibinfo {author} {\bibfnamefont {J.}~\bibnamefont {Li}}, \bibinfo {author} {\bibfnamefont {Z.}~\bibnamefont {Su}},\ and\ \bibinfo {author} {\bibfnamefont {J.}~\bibnamefont {Zhu}},\ }\bibfield  {title} {\bibinfo {title} {Transient non-hermitian skin effect},\ }\href {https://doi.org/10.1038/s41467-022-35448-2} {\bibfield  {journal} {\bibinfo  {journal} {Nat. Commun.}\ }\textbf {\bibinfo {volume} {13}},\ \bibinfo {pages} {7668} (\bibinfo {year} {2022})}\BibitemShut {NoStop}%
\bibitem [{\citenamefont {Liang}\ \emph {et~al.}(2022)\citenamefont {Liang}, \citenamefont {Xie}, \citenamefont {Dong}, \citenamefont {Li}, \citenamefont {Li}, \citenamefont {Gadway}, \citenamefont {Yi},\ and\ \citenamefont {Yan}}]{PhysRevLett.129.070401}%
  \BibitemOpen
  \bibfield  {author} {\bibinfo {author} {\bibfnamefont {Q.}~\bibnamefont {Liang}}, \bibinfo {author} {\bibfnamefont {D.}~\bibnamefont {Xie}}, \bibinfo {author} {\bibfnamefont {Z.}~\bibnamefont {Dong}}, \bibinfo {author} {\bibfnamefont {H.}~\bibnamefont {Li}}, \bibinfo {author} {\bibfnamefont {H.}~\bibnamefont {Li}}, \bibinfo {author} {\bibfnamefont {B.}~\bibnamefont {Gadway}}, \bibinfo {author} {\bibfnamefont {W.}~\bibnamefont {Yi}},\ and\ \bibinfo {author} {\bibfnamefont {B.}~\bibnamefont {Yan}},\ }\bibfield  {title} {\bibinfo {title} {Dynamic signatures of non-hermitian skin effect and topology in ultracold atoms},\ }\href {https://doi.org/10.1103/PhysRevLett.129.070401} {\bibfield  {journal} {\bibinfo  {journal} {Phys. Rev. Lett.}\ }\textbf {\bibinfo {volume} {129}},\ \bibinfo {pages} {070401} (\bibinfo {year} {2022})}\BibitemShut {NoStop}%
\bibitem [{\citenamefont {Lin}\ \emph {et~al.}(2023)\citenamefont {Lin}, \citenamefont {Tai}, \citenamefont {Li},\ and\ \citenamefont {Lee}}]{lin2023topological}%
  \BibitemOpen
  \bibfield  {author} {\bibinfo {author} {\bibfnamefont {R.}~\bibnamefont {Lin}}, \bibinfo {author} {\bibfnamefont {T.}~\bibnamefont {Tai}}, \bibinfo {author} {\bibfnamefont {L.}~\bibnamefont {Li}},\ and\ \bibinfo {author} {\bibfnamefont {C.~H.}\ \bibnamefont {Lee}},\ }\bibfield  {title} {\bibinfo {title} {Topological non-hermitian skin effect},\ }\href {https://doi.org/10.1007/s11467-023-1309-z} {\bibfield  {journal} {\bibinfo  {journal} {Front. Phys.}\ }\textbf {\bibinfo {volume} {18}},\ \bibinfo {pages} {53605} (\bibinfo {year} {2023})}\BibitemShut {NoStop}%
\bibitem [{\citenamefont {Kawabata}\ \emph {et~al.}(2023)\citenamefont {Kawabata}, \citenamefont {Numasawa},\ and\ \citenamefont {Ryu}}]{PhysRevX.13.021007}%
  \BibitemOpen
  \bibfield  {author} {\bibinfo {author} {\bibfnamefont {K.}~\bibnamefont {Kawabata}}, \bibinfo {author} {\bibfnamefont {T.}~\bibnamefont {Numasawa}},\ and\ \bibinfo {author} {\bibfnamefont {S.}~\bibnamefont {Ryu}},\ }\bibfield  {title} {\bibinfo {title} {Entanglement phase transition induced by the non-hermitian skin effect},\ }\href {https://doi.org/10.1103/PhysRevX.13.021007} {\bibfield  {journal} {\bibinfo  {journal} {Phys. Rev. X}\ }\textbf {\bibinfo {volume} {13}},\ \bibinfo {pages} {021007} (\bibinfo {year} {2023})}\BibitemShut {NoStop}%
\bibitem [{\citenamefont {Li}\ \emph {et~al.}(2024)\citenamefont {Li}, \citenamefont {Wang}, \citenamefont {Wang}, \citenamefont {Lin}, \citenamefont {Ma},\ and\ \citenamefont {Jiang}}]{li2024observation}%
  \BibitemOpen
  \bibfield  {author} {\bibinfo {author} {\bibfnamefont {Z.}~\bibnamefont {Li}}, \bibinfo {author} {\bibfnamefont {L.-W.}\ \bibnamefont {Wang}}, \bibinfo {author} {\bibfnamefont {X.}~\bibnamefont {Wang}}, \bibinfo {author} {\bibfnamefont {Z.-K.}\ \bibnamefont {Lin}}, \bibinfo {author} {\bibfnamefont {G.}~\bibnamefont {Ma}},\ and\ \bibinfo {author} {\bibfnamefont {J.-H.}\ \bibnamefont {Jiang}},\ }\bibfield  {title} {\bibinfo {title} {Observation of dynamic non-hermitian skin effects},\ }\href {https://doi.org/10.1038/s41467-024-50776-1} {\bibfield  {journal} {\bibinfo  {journal} {Nat. Commun.}\ }\textbf {\bibinfo {volume} {15}},\ \bibinfo {pages} {6544} (\bibinfo {year} {2024})}\BibitemShut {NoStop}%
\bibitem [{\citenamefont {El-Ganainy}\ \emph {et~al.}(2018)\citenamefont {El-Ganainy}, \citenamefont {Makris}, \citenamefont {Khajavikhan}, \citenamefont {Musslimani}, \citenamefont {Rotter},\ and\ \citenamefont {Christodoulides}}]{el2018non}%
  \BibitemOpen
  \bibfield  {author} {\bibinfo {author} {\bibfnamefont {R.}~\bibnamefont {El-Ganainy}}, \bibinfo {author} {\bibfnamefont {K.~G.}\ \bibnamefont {Makris}}, \bibinfo {author} {\bibfnamefont {M.}~\bibnamefont {Khajavikhan}}, \bibinfo {author} {\bibfnamefont {Z.~H.}\ \bibnamefont {Musslimani}}, \bibinfo {author} {\bibfnamefont {S.}~\bibnamefont {Rotter}},\ and\ \bibinfo {author} {\bibfnamefont {D.~N.}\ \bibnamefont {Christodoulides}},\ }\bibfield  {title} {\bibinfo {title} {Non-hermitian physics and pt symmetry},\ }\href {https://doi.org/10.1038/nphys4323} {\bibfield  {journal} {\bibinfo  {journal} {Nat. Phys.}\ }\textbf {\bibinfo {volume} {14}},\ \bibinfo {pages} {11} (\bibinfo {year} {2018})}\BibitemShut {NoStop}%
\bibitem [{\citenamefont {Yao}\ \emph {et~al.}(2018)\citenamefont {Yao}, \citenamefont {Song},\ and\ \citenamefont {Wang}}]{PhysRevLett.121.136802}%
  \BibitemOpen
  \bibfield  {author} {\bibinfo {author} {\bibfnamefont {S.}~\bibnamefont {Yao}}, \bibinfo {author} {\bibfnamefont {F.}~\bibnamefont {Song}},\ and\ \bibinfo {author} {\bibfnamefont {Z.}~\bibnamefont {Wang}},\ }\bibfield  {title} {\bibinfo {title} {{Non-Hermitian Chern Bands}},\ }\href {https://doi.org/10.1103/PhysRevLett.121.136802} {\bibfield  {journal} {\bibinfo  {journal} {Phys. Rev. Lett.}\ }\textbf {\bibinfo {volume} {121}},\ \bibinfo {pages} {136802} (\bibinfo {year} {2018})}\BibitemShut {NoStop}%
\bibitem [{\citenamefont {Yokomizo}\ and\ \citenamefont {Murakami}(2019)}]{PhysRevLett.123.066404}%
  \BibitemOpen
  \bibfield  {author} {\bibinfo {author} {\bibfnamefont {K.}~\bibnamefont {Yokomizo}}\ and\ \bibinfo {author} {\bibfnamefont {S.}~\bibnamefont {Murakami}},\ }\bibfield  {title} {\bibinfo {title} {{Non-Bloch Band Theory of Non-Hermitian Systems}},\ }\href {https://doi.org/10.1103/PhysRevLett.123.066404} {\bibfield  {journal} {\bibinfo  {journal} {Phys. Rev. Lett.}\ }\textbf {\bibinfo {volume} {123}},\ \bibinfo {pages} {066404} (\bibinfo {year} {2019})}\BibitemShut {NoStop}%
\bibitem [{\citenamefont {Pan}\ \emph {et~al.}(2020)\citenamefont {Pan}, \citenamefont {Chen}, \citenamefont {Chen},\ and\ \citenamefont {Zhai}}]{pan2020non}%
  \BibitemOpen
  \bibfield  {author} {\bibinfo {author} {\bibfnamefont {L.}~\bibnamefont {Pan}}, \bibinfo {author} {\bibfnamefont {X.}~\bibnamefont {Chen}}, \bibinfo {author} {\bibfnamefont {Y.}~\bibnamefont {Chen}},\ and\ \bibinfo {author} {\bibfnamefont {H.}~\bibnamefont {Zhai}},\ }\bibfield  {title} {\bibinfo {title} {{Non-Hermitian linear response theory}},\ }\href {https://doi.org/10.1038/s41567-020-0889-6} {\bibfield  {journal} {\bibinfo  {journal} {Nat. Phys.}\ }\textbf {\bibinfo {volume} {16}},\ \bibinfo {pages} {767} (\bibinfo {year} {2020})}\BibitemShut {NoStop}%
\bibitem [{\citenamefont {Song}\ \emph {et~al.}(2022)\citenamefont {Song}, \citenamefont {Chen}, \citenamefont {Xiong},\ and\ \citenamefont {Wang}}]{Song:22}%
  \BibitemOpen
  \bibfield  {author} {\bibinfo {author} {\bibfnamefont {Y.}~\bibnamefont {Song}}, \bibinfo {author} {\bibfnamefont {Y.}~\bibnamefont {Chen}}, \bibinfo {author} {\bibfnamefont {W.}~\bibnamefont {Xiong}},\ and\ \bibinfo {author} {\bibfnamefont {M.}~\bibnamefont {Wang}},\ }\bibfield  {title} {\bibinfo {title} {{Flexible light manipulation in non-Hermitian frequency Su--Schrieffer--Heeger lattice}},\ }\href {https://doi.org/10.1364/OL.452983} {\bibfield  {journal} {\bibinfo  {journal} {Opt. Lett.}\ }\textbf {\bibinfo {volume} {47}},\ \bibinfo {pages} {1646} (\bibinfo {year} {2022})}\BibitemShut {NoStop}%
\bibitem [{\citenamefont {Li}\ \emph {et~al.}(2020{\natexlab{b}})\citenamefont {Li}, \citenamefont {Lee},\ and\ \citenamefont {Gong}}]{PhysRevLett.124.250402}%
  \BibitemOpen
  \bibfield  {author} {\bibinfo {author} {\bibfnamefont {L.}~\bibnamefont {Li}}, \bibinfo {author} {\bibfnamefont {C.~H.}\ \bibnamefont {Lee}},\ and\ \bibinfo {author} {\bibfnamefont {J.}~\bibnamefont {Gong}},\ }\bibfield  {title} {\bibinfo {title} {{Topological Switch for Non-Hermitian Skin Effect in Cold-Atom Systems with Loss}},\ }\href {https://doi.org/10.1103/PhysRevLett.124.250402} {\bibfield  {journal} {\bibinfo  {journal} {Phys. Rev. Lett.}\ }\textbf {\bibinfo {volume} {124}},\ \bibinfo {pages} {250402} (\bibinfo {year} {2020}{\natexlab{b}})}\BibitemShut {NoStop}%
\bibitem [{\citenamefont {Qin}\ \emph {et~al.}(2024)\citenamefont {Qin}, \citenamefont {Shen}, \citenamefont {Li},\ and\ \citenamefont {Lee}}]{PhysRevA.109.053311}%
  \BibitemOpen
  \bibfield  {author} {\bibinfo {author} {\bibfnamefont {F.}~\bibnamefont {Qin}}, \bibinfo {author} {\bibfnamefont {R.}~\bibnamefont {Shen}}, \bibinfo {author} {\bibfnamefont {L.}~\bibnamefont {Li}},\ and\ \bibinfo {author} {\bibfnamefont {C.~H.}\ \bibnamefont {Lee}},\ }\bibfield  {title} {\bibinfo {title} {{Kinked linear response from non-Hermitian cold-atom pumping}},\ }\href {https://doi.org/10.1103/PhysRevA.109.053311} {\bibfield  {journal} {\bibinfo  {journal} {Phys. Rev. A}\ }\textbf {\bibinfo {volume} {109}},\ \bibinfo {pages} {053311} (\bibinfo {year} {2024})}\BibitemShut {NoStop}%
\bibitem [{\citenamefont {Luo}\ \emph {et~al.}(2024)\citenamefont {Luo}, \citenamefont {Liu},\ and\ \citenamefont {Liang}}]{PhysRevA.110.063320}%
  \BibitemOpen
  \bibfield  {author} {\bibinfo {author} {\bibfnamefont {Y.}~\bibnamefont {Luo}}, \bibinfo {author} {\bibfnamefont {N.}~\bibnamefont {Liu}},\ and\ \bibinfo {author} {\bibfnamefont {J.-Q.}\ \bibnamefont {Liang}},\ }\bibfield  {title} {\bibinfo {title} {{Quantum phase transition and exceptional points of a non-Hermitian Hamiltonian for cold atoms in a dissipative optical cavity with nonlinear atom-photon interactions}},\ }\href {https://doi.org/10.1103/PhysRevA.110.063320} {\bibfield  {journal} {\bibinfo  {journal} {Phys. Rev. A}\ }\textbf {\bibinfo {volume} {110}},\ \bibinfo {pages} {063320} (\bibinfo {year} {2024})}\BibitemShut {NoStop}%
\bibitem [{\citenamefont {Liu}\ \emph {et~al.}(2024)\citenamefont {Liu}, \citenamefont {Ren}, \citenamefont {Wong}, \citenamefont {Zhao}, \citenamefont {He}, \citenamefont {Pak}, \citenamefont {Jo},\ and\ \citenamefont {Li}}]{PhysRevA.109.053305}%
  \BibitemOpen
  \bibfield  {author} {\bibinfo {author} {\bibfnamefont {D.}~\bibnamefont {Liu}}, \bibinfo {author} {\bibfnamefont {Z.}~\bibnamefont {Ren}}, \bibinfo {author} {\bibfnamefont {W.~C.}\ \bibnamefont {Wong}}, \bibinfo {author} {\bibfnamefont {E.}~\bibnamefont {Zhao}}, \bibinfo {author} {\bibfnamefont {C.}~\bibnamefont {He}}, \bibinfo {author} {\bibfnamefont {K.~K.}\ \bibnamefont {Pak}}, \bibinfo {author} {\bibfnamefont {G.-B.}\ \bibnamefont {Jo}},\ and\ \bibinfo {author} {\bibfnamefont {J.}~\bibnamefont {Li}},\ }\bibfield  {title} {\bibinfo {title} {{Complete interband transitions for non-Hermitian spin-orbit-coupled cold-atom systems}},\ }\href {https://doi.org/10.1103/PhysRevA.109.053305} {\bibfield  {journal} {\bibinfo  {journal} {Phys. Rev. A}\ }\textbf {\bibinfo {volume} {109}},\ \bibinfo {pages} {053305} (\bibinfo {year} {2024})}\BibitemShut {NoStop}%
\bibitem [{\citenamefont {Zhao}\ \emph {et~al.}(2019)\citenamefont {Zhao}, \citenamefont {Qiao}, \citenamefont {Wu}, \citenamefont {Midya}, \citenamefont {Longhi},\ and\ \citenamefont {Feng}}]{doi:10.1126/science.aay1064}%
  \BibitemOpen
  \bibfield  {author} {\bibinfo {author} {\bibfnamefont {H.}~\bibnamefont {Zhao}}, \bibinfo {author} {\bibfnamefont {X.}~\bibnamefont {Qiao}}, \bibinfo {author} {\bibfnamefont {T.}~\bibnamefont {Wu}}, \bibinfo {author} {\bibfnamefont {B.}~\bibnamefont {Midya}}, \bibinfo {author} {\bibfnamefont {S.}~\bibnamefont {Longhi}},\ and\ \bibinfo {author} {\bibfnamefont {L.}~\bibnamefont {Feng}},\ }\bibfield  {title} {\bibinfo {title} {{Non-Hermitian topological light steering}},\ }\href {https://doi.org/10.1126/science.aay1064} {\bibfield  {journal} {\bibinfo  {journal} {Science}\ }\textbf {\bibinfo {volume} {365}},\ \bibinfo {pages} {1163} (\bibinfo {year} {2019})}\BibitemShut {NoStop}%
\bibitem [{\citenamefont {Longhi}(2019{\natexlab{b}})}]{Longhi:19}%
  \BibitemOpen
  \bibfield  {author} {\bibinfo {author} {\bibfnamefont {S.}~\bibnamefont {Longhi}},\ }\bibfield  {title} {\bibinfo {title} {{Non-Bloch $\mathcal{PT}$ symmetry breaking in non-Hermitian photonic quantum walks}},\ }\href {https://doi.org/10.1364/OL.44.005804} {\bibfield  {journal} {\bibinfo  {journal} {Opt. Lett.}\ }\textbf {\bibinfo {volume} {44}},\ \bibinfo {pages} {5804} (\bibinfo {year} {2019}{\natexlab{b}})}\BibitemShut {NoStop}%
\bibitem [{\citenamefont {Weidemann}\ \emph {et~al.}(2020)\citenamefont {Weidemann}, \citenamefont {Kremer}, \citenamefont {Helbig}, \citenamefont {Hofmann}, \citenamefont {Stegmaier}, \citenamefont {Greiter}, \citenamefont {Thomale},\ and\ \citenamefont {Szameit}}]{Weidemann2020topological}%
  \BibitemOpen
  \bibfield  {author} {\bibinfo {author} {\bibfnamefont {S.}~\bibnamefont {Weidemann}}, \bibinfo {author} {\bibfnamefont {M.}~\bibnamefont {Kremer}}, \bibinfo {author} {\bibfnamefont {T.}~\bibnamefont {Helbig}}, \bibinfo {author} {\bibfnamefont {T.}~\bibnamefont {Hofmann}}, \bibinfo {author} {\bibfnamefont {A.}~\bibnamefont {Stegmaier}}, \bibinfo {author} {\bibfnamefont {M.}~\bibnamefont {Greiter}}, \bibinfo {author} {\bibfnamefont {R.}~\bibnamefont {Thomale}},\ and\ \bibinfo {author} {\bibfnamefont {A.}~\bibnamefont {Szameit}},\ }\bibfield  {title} {\bibinfo {title} {{Topological funneling of light}},\ }\href {https://www.science.org/doi/abs/10.1126/science.aaz8727} {\bibfield  {journal} {\bibinfo  {journal} {Science}\ }\textbf {\bibinfo {volume} {368}},\ \bibinfo {pages} {311} (\bibinfo {year} {2020})}\BibitemShut {NoStop}%
\bibitem [{\citenamefont {Li}\ \emph {et~al.}(2020{\natexlab{c}})\citenamefont {Li}, \citenamefont {Zhirihin}, \citenamefont {Gorlach}, \citenamefont {Ni}, \citenamefont {Filonov}, \citenamefont {Slobozhanyuk}, \citenamefont {Al{\`u}},\ and\ \citenamefont {Khanikaev}}]{li2020higher}%
  \BibitemOpen
  \bibfield  {author} {\bibinfo {author} {\bibfnamefont {M.}~\bibnamefont {Li}}, \bibinfo {author} {\bibfnamefont {D.}~\bibnamefont {Zhirihin}}, \bibinfo {author} {\bibfnamefont {M.}~\bibnamefont {Gorlach}}, \bibinfo {author} {\bibfnamefont {X.}~\bibnamefont {Ni}}, \bibinfo {author} {\bibfnamefont {D.}~\bibnamefont {Filonov}}, \bibinfo {author} {\bibfnamefont {A.}~\bibnamefont {Slobozhanyuk}}, \bibinfo {author} {\bibfnamefont {A.}~\bibnamefont {Al{\`u}}},\ and\ \bibinfo {author} {\bibfnamefont {A.~B.}\ \bibnamefont {Khanikaev}},\ }\bibfield  {title} {\bibinfo {title} {{Higher-order topological states in photonic kagome crystals with long-range interactions}},\ }\href {https://doi.org/10.1038/s41566-019-0561-9} {\bibfield  {journal} {\bibinfo  {journal} {Nat. Photonics}\ }\textbf {\bibinfo {volume} {14}},\ \bibinfo {pages} {89} (\bibinfo {year} {2020}{\natexlab{c}})}\BibitemShut {NoStop}%
\bibitem [{\citenamefont {Li}\ \emph {et~al.}(2021)\citenamefont {Li}, \citenamefont {Liu}, \citenamefont {Lin}, \citenamefont {Ng},\ and\ \citenamefont {Chan}}]{li2021non}%
  \BibitemOpen
  \bibfield  {author} {\bibinfo {author} {\bibfnamefont {X.}~\bibnamefont {Li}}, \bibinfo {author} {\bibfnamefont {Y.}~\bibnamefont {Liu}}, \bibinfo {author} {\bibfnamefont {Z.}~\bibnamefont {Lin}}, \bibinfo {author} {\bibfnamefont {J.}~\bibnamefont {Ng}},\ and\ \bibinfo {author} {\bibfnamefont {C.~T.}\ \bibnamefont {Chan}},\ }\bibfield  {title} {\bibinfo {title} {{Non-Hermitian physics for optical manipulation uncovers inherent instability of large clusters}},\ }\href {https://doi.org/10.1038/s41467-021-26732-8} {\bibfield  {journal} {\bibinfo  {journal} {Nat. Commun.}\ }\textbf {\bibinfo {volume} {12}},\ \bibinfo {pages} {6597} (\bibinfo {year} {2021})}\BibitemShut {NoStop}%
\bibitem [{\citenamefont {Wang}\ \emph {et~al.}(2023)\citenamefont {Wang}, \citenamefont {Qian},\ and\ \citenamefont {Jiang}}]{Wang:23}%
  \BibitemOpen
  \bibfield  {author} {\bibinfo {author} {\bibfnamefont {Q.}~\bibnamefont {Wang}}, \bibinfo {author} {\bibfnamefont {J.}~\bibnamefont {Qian}},\ and\ \bibinfo {author} {\bibfnamefont {L.}~\bibnamefont {Jiang}},\ }\bibfield  {title} {\bibinfo {title} {{Non-Hermitian kagome photonic crystal with a totally topological spatial mode selection}},\ }\href {https://doi.org/10.1364/OE.482836} {\bibfield  {journal} {\bibinfo  {journal} {Opt. Express}\ }\textbf {\bibinfo {volume} {31}},\ \bibinfo {pages} {5363} (\bibinfo {year} {2023})}\BibitemShut {NoStop}%
\bibitem [{\citenamefont {Zhu}\ and\ \citenamefont {Gong}(2023)}]{PhysRevB.108.035406}%
  \BibitemOpen
  \bibfield  {author} {\bibinfo {author} {\bibfnamefont {W.}~\bibnamefont {Zhu}}\ and\ \bibinfo {author} {\bibfnamefont {J.}~\bibnamefont {Gong}},\ }\bibfield  {title} {\bibinfo {title} {{Photonic corner skin modes in non-Hermitian photonic crystals}},\ }\href {https://doi.org/10.1103/PhysRevB.108.035406} {\bibfield  {journal} {\bibinfo  {journal} {Phys. Rev. B}\ }\textbf {\bibinfo {volume} {108}},\ \bibinfo {pages} {035406} (\bibinfo {year} {2023})}\BibitemShut {NoStop}%
\bibitem [{\citenamefont {Reisenbauer}\ \emph {et~al.}(2024)\citenamefont {Reisenbauer}, \citenamefont {Rudolph}, \citenamefont {Egyed}, \citenamefont {Hornberger}, \citenamefont {Zasedatelev}, \citenamefont {Abuzarli}, \citenamefont {Stickler},\ and\ \citenamefont {Deli{\'c}}}]{reisenbauer2024non}%
  \BibitemOpen
  \bibfield  {author} {\bibinfo {author} {\bibfnamefont {M.}~\bibnamefont {Reisenbauer}}, \bibinfo {author} {\bibfnamefont {H.}~\bibnamefont {Rudolph}}, \bibinfo {author} {\bibfnamefont {L.}~\bibnamefont {Egyed}}, \bibinfo {author} {\bibfnamefont {K.}~\bibnamefont {Hornberger}}, \bibinfo {author} {\bibfnamefont {A.~V.}\ \bibnamefont {Zasedatelev}}, \bibinfo {author} {\bibfnamefont {M.}~\bibnamefont {Abuzarli}}, \bibinfo {author} {\bibfnamefont {B.~A.}\ \bibnamefont {Stickler}},\ and\ \bibinfo {author} {\bibfnamefont {U.}~\bibnamefont {Deli{\'c}}},\ }\bibfield  {title} {\bibinfo {title} {{Non-Hermitian dynamics and non-reciprocity of optically coupled nanoparticles}},\ }\href {https://doi.org/10.1038/s41567-024-02589-8} {\bibfield  {journal} {\bibinfo  {journal} {Nat. Phys.}\ }\textbf {\bibinfo {volume} {20}},\ \bibinfo {pages} {1629} (\bibinfo {year} {2024})}\BibitemShut {NoStop}%
\bibitem [{\citenamefont {Meng}\ \emph {et~al.}(2024)\citenamefont {Meng}, \citenamefont {Ang},\ and\ \citenamefont {Lee}}]{10.1063/5.0183826}%
  \BibitemOpen
  \bibfield  {author} {\bibinfo {author} {\bibfnamefont {H.}~\bibnamefont {Meng}}, \bibinfo {author} {\bibfnamefont {Y.~S.}\ \bibnamefont {Ang}},\ and\ \bibinfo {author} {\bibfnamefont {C.~H.}\ \bibnamefont {Lee}},\ }\bibfield  {title} {\bibinfo {title} {{Exceptional points in non-Hermitian systems: Applications and recent developments}},\ }\href {https://doi.org/10.1063/5.0183826} {\bibfield  {journal} {\bibinfo  {journal} {Appl. Phys. Lett.}\ }\textbf {\bibinfo {volume} {124}},\ \bibinfo {pages} {060502} (\bibinfo {year} {2024})}\BibitemShut {NoStop}%
\bibitem [{\citenamefont {Zheng}\ \emph {et~al.}(2024)\citenamefont {Zheng}, \citenamefont {Jalali~Mehrabad}, \citenamefont {Vannucci}, \citenamefont {Li}, \citenamefont {Dutt}, \citenamefont {Hafezi}, \citenamefont {Mittal},\ and\ \citenamefont {Waks}}]{Zheng2024}%
  \BibitemOpen
  \bibfield  {author} {\bibinfo {author} {\bibfnamefont {X.}~\bibnamefont {Zheng}}, \bibinfo {author} {\bibfnamefont {M.}~\bibnamefont {Jalali~Mehrabad}}, \bibinfo {author} {\bibfnamefont {J.}~\bibnamefont {Vannucci}}, \bibinfo {author} {\bibfnamefont {K.}~\bibnamefont {Li}}, \bibinfo {author} {\bibfnamefont {A.}~\bibnamefont {Dutt}}, \bibinfo {author} {\bibfnamefont {M.}~\bibnamefont {Hafezi}}, \bibinfo {author} {\bibfnamefont {S.}~\bibnamefont {Mittal}},\ and\ \bibinfo {author} {\bibfnamefont {E.}~\bibnamefont {Waks}},\ }\bibfield  {title} {\bibinfo {title} {Dynamic control of 2d non-hermitian photonic corner skin modes in synthetic dimensions},\ }\href {https://doi.org/10.1038/s41467-024-55236-4} {\bibfield  {journal} {\bibinfo  {journal} {Nat. Commun.}\ }\textbf {\bibinfo {volume} {15}},\ \bibinfo {pages} {10881} (\bibinfo {year} {2024})}\BibitemShut {NoStop}%
\bibitem [{\citenamefont {Roccati}\ \emph {et~al.}(2024)\citenamefont {Roccati}, \citenamefont {Bello}, \citenamefont {Gong}, \citenamefont {Ueda}, \citenamefont {Ciccarello}, \citenamefont {Chenu},\ and\ \citenamefont {Carollo}}]{Roccati2024}%
  \BibitemOpen
  \bibfield  {author} {\bibinfo {author} {\bibfnamefont {F.}~\bibnamefont {Roccati}}, \bibinfo {author} {\bibfnamefont {M.}~\bibnamefont {Bello}}, \bibinfo {author} {\bibfnamefont {Z.}~\bibnamefont {Gong}}, \bibinfo {author} {\bibfnamefont {M.}~\bibnamefont {Ueda}}, \bibinfo {author} {\bibfnamefont {F.}~\bibnamefont {Ciccarello}}, \bibinfo {author} {\bibfnamefont {A.}~\bibnamefont {Chenu}},\ and\ \bibinfo {author} {\bibfnamefont {A.}~\bibnamefont {Carollo}},\ }\bibfield  {title} {\bibinfo {title} {Hermitian and non-hermitian topology from photon-mediated interactions},\ }\href {https://doi.org/10.1038/s41467-024-46471-w} {\bibfield  {journal} {\bibinfo  {journal} {Nat. Commun.}\ }\textbf {\bibinfo {volume} {15}},\ \bibinfo {pages} {2400} (\bibinfo {year} {2024})}\BibitemShut {NoStop}%
\bibitem [{\citenamefont {Gao}\ \emph {et~al.}(2024)\citenamefont {Gao}, \citenamefont {Sheng}, \citenamefont {Zhao}, \citenamefont {He}, \citenamefont {Lu}, \citenamefont {Chen}, \citenamefont {Ding}, \citenamefont {Zhu},\ and\ \citenamefont {Liu}}]{PhysRevB.110.094308}%
  \BibitemOpen
  \bibfield  {author} {\bibinfo {author} {\bibfnamefont {M.}~\bibnamefont {Gao}}, \bibinfo {author} {\bibfnamefont {C.}~\bibnamefont {Sheng}}, \bibinfo {author} {\bibfnamefont {Y.}~\bibnamefont {Zhao}}, \bibinfo {author} {\bibfnamefont {R.}~\bibnamefont {He}}, \bibinfo {author} {\bibfnamefont {L.}~\bibnamefont {Lu}}, \bibinfo {author} {\bibfnamefont {W.}~\bibnamefont {Chen}}, \bibinfo {author} {\bibfnamefont {K.}~\bibnamefont {Ding}}, \bibinfo {author} {\bibfnamefont {S.}~\bibnamefont {Zhu}},\ and\ \bibinfo {author} {\bibfnamefont {H.}~\bibnamefont {Liu}},\ }\bibfield  {title} {\bibinfo {title} {Quantum walks of correlated photons in non-hermitian photonic lattices},\ }\href {https://doi.org/10.1103/PhysRevB.110.094308} {\bibfield  {journal} {\bibinfo  {journal} {Phys. Rev. B}\ }\textbf {\bibinfo {volume} {110}},\ \bibinfo {pages} {094308} (\bibinfo {year} {2024})}\BibitemShut {NoStop}%
\bibitem [{\citenamefont {Zhang}\ \emph {et~al.}(2024{\natexlab{a}})\citenamefont {Zhang}, \citenamefont {Dai},\ and\ \citenamefont {Xiang}}]{PhysRevB.110.104103}%
  \BibitemOpen
  \bibfield  {author} {\bibinfo {author} {\bibfnamefont {Y.}~\bibnamefont {Zhang}}, \bibinfo {author} {\bibfnamefont {X.}~\bibnamefont {Dai}},\ and\ \bibinfo {author} {\bibfnamefont {Y.}~\bibnamefont {Xiang}},\ }\bibfield  {title} {\bibinfo {title} {Topological hierarchy in non-hermitian three-dimensional photonic crystals},\ }\href {https://doi.org/10.1103/PhysRevB.110.104103} {\bibfield  {journal} {\bibinfo  {journal} {Phys. Rev. B}\ }\textbf {\bibinfo {volume} {110}},\ \bibinfo {pages} {104103} (\bibinfo {year} {2024}{\natexlab{a}})}\BibitemShut {NoStop}%
\bibitem [{\citenamefont {Feng}\ \emph {et~al.}(2025)\citenamefont {Feng}, \citenamefont {Wu}, \citenamefont {Gao}, \citenamefont {Zhao}, \citenamefont {Wu}, \citenamefont {Zhang}, \citenamefont {Ge},\ and\ \citenamefont {Feng}}]{Feng2025}%
  \BibitemOpen
  \bibfield  {author} {\bibinfo {author} {\bibfnamefont {X.}~\bibnamefont {Feng}}, \bibinfo {author} {\bibfnamefont {T.}~\bibnamefont {Wu}}, \bibinfo {author} {\bibfnamefont {Z.}~\bibnamefont {Gao}}, \bibinfo {author} {\bibfnamefont {H.}~\bibnamefont {Zhao}}, \bibinfo {author} {\bibfnamefont {S.}~\bibnamefont {Wu}}, \bibinfo {author} {\bibfnamefont {Y.}~\bibnamefont {Zhang}}, \bibinfo {author} {\bibfnamefont {L.}~\bibnamefont {Ge}},\ and\ \bibinfo {author} {\bibfnamefont {L.}~\bibnamefont {Feng}},\ }\bibfield  {title} {\bibinfo {title} {Non-hermitian hybrid silicon photonic switching},\ }\href {https://doi.org/10.1038/s41566-024-01579-9} {\bibfield  {journal} {\bibinfo  {journal} {Nat. Photonics}\ }\textbf {\bibinfo {volume} {19}},\ \bibinfo {pages} {264} (\bibinfo {year} {2025})}\BibitemShut {NoStop}%
\bibitem [{\citenamefont {Zhang}\ \emph {et~al.}(2019{\natexlab{a}})\citenamefont {Zhang}, \citenamefont {Rosendo~L\'opez}, \citenamefont {Cheng}, \citenamefont {Liu},\ and\ \citenamefont {Christensen}}]{PhysRevLett.122.195501}%
  \BibitemOpen
  \bibfield  {author} {\bibinfo {author} {\bibfnamefont {Z.}~\bibnamefont {Zhang}}, \bibinfo {author} {\bibfnamefont {M.}~\bibnamefont {Rosendo~L\'opez}}, \bibinfo {author} {\bibfnamefont {Y.}~\bibnamefont {Cheng}}, \bibinfo {author} {\bibfnamefont {X.}~\bibnamefont {Liu}},\ and\ \bibinfo {author} {\bibfnamefont {J.}~\bibnamefont {Christensen}},\ }\bibfield  {title} {\bibinfo {title} {{Non-Hermitian Sonic Second-Order Topological Insulator}},\ }\href {https://doi.org/10.1103/PhysRevLett.122.195501} {\bibfield  {journal} {\bibinfo  {journal} {Phys. Rev. Lett.}\ }\textbf {\bibinfo {volume} {122}},\ \bibinfo {pages} {195501} (\bibinfo {year} {2019}{\natexlab{a}})}\BibitemShut {NoStop}%
\bibitem [{\citenamefont {Rosendo~L{\'o}pez}\ \emph {et~al.}(2019)\citenamefont {Rosendo~L{\'o}pez}, \citenamefont {Zhang}, \citenamefont {Torrent},\ and\ \citenamefont {Christensen}}]{rosendo2019multiple}%
  \BibitemOpen
  \bibfield  {author} {\bibinfo {author} {\bibfnamefont {M.}~\bibnamefont {Rosendo~L{\'o}pez}}, \bibinfo {author} {\bibfnamefont {Z.}~\bibnamefont {Zhang}}, \bibinfo {author} {\bibfnamefont {D.}~\bibnamefont {Torrent}},\ and\ \bibinfo {author} {\bibfnamefont {J.}~\bibnamefont {Christensen}},\ }\bibfield  {title} {\bibinfo {title} {{Multiple scattering theory of non-Hermitian sonic second-order topological insulators}},\ }\href {https://doi.org/10.1038/s42005-019-0233-6} {\bibfield  {journal} {\bibinfo  {journal} {Commun. Phys.}\ }\textbf {\bibinfo {volume} {2}},\ \bibinfo {pages} {132} (\bibinfo {year} {2019})}\BibitemShut {NoStop}%
\bibitem [{\citenamefont {Gao}\ \emph {et~al.}(2021)\citenamefont {Gao}, \citenamefont {Xue}, \citenamefont {Gu}, \citenamefont {Liu}, \citenamefont {Zhu},\ and\ \citenamefont {Zhang}}]{gao2021non}%
  \BibitemOpen
  \bibfield  {author} {\bibinfo {author} {\bibfnamefont {H.}~\bibnamefont {Gao}}, \bibinfo {author} {\bibfnamefont {H.}~\bibnamefont {Xue}}, \bibinfo {author} {\bibfnamefont {Z.}~\bibnamefont {Gu}}, \bibinfo {author} {\bibfnamefont {T.}~\bibnamefont {Liu}}, \bibinfo {author} {\bibfnamefont {J.}~\bibnamefont {Zhu}},\ and\ \bibinfo {author} {\bibfnamefont {B.}~\bibnamefont {Zhang}},\ }\bibfield  {title} {\bibinfo {title} {{Non-Hermitian route to higher-order topology in an acoustic crystal}},\ }\href {https://doi.org/10.1038/s41467-021-22223-y} {\bibfield  {journal} {\bibinfo  {journal} {Nat. Commun.}\ }\textbf {\bibinfo {volume} {12}},\ \bibinfo {pages} {1888} (\bibinfo {year} {2021})}\BibitemShut {NoStop}%
\bibitem [{\citenamefont {Gu}\ \emph {et~al.}(2021)\citenamefont {Gu}, \citenamefont {Gao}, \citenamefont {Cao}, \citenamefont {Liu}, \citenamefont {Zhu},\ and\ \citenamefont {Zhu}}]{PhysRevApplied.16.057001}%
  \BibitemOpen
  \bibfield  {author} {\bibinfo {author} {\bibfnamefont {Z.}~\bibnamefont {Gu}}, \bibinfo {author} {\bibfnamefont {H.}~\bibnamefont {Gao}}, \bibinfo {author} {\bibfnamefont {P.-C.}\ \bibnamefont {Cao}}, \bibinfo {author} {\bibfnamefont {T.}~\bibnamefont {Liu}}, \bibinfo {author} {\bibfnamefont {X.-F.}\ \bibnamefont {Zhu}},\ and\ \bibinfo {author} {\bibfnamefont {J.}~\bibnamefont {Zhu}},\ }\bibfield  {title} {\bibinfo {title} {{Controlling Sound in Non-Hermitian Acoustic Systems}},\ }\href {https://doi.org/10.1103/PhysRevApplied.16.057001} {\bibfield  {journal} {\bibinfo  {journal} {Phys. Rev. Appl.}\ }\textbf {\bibinfo {volume} {16}},\ \bibinfo {pages} {057001} (\bibinfo {year} {2021})}\BibitemShut {NoStop}%
\bibitem [{\citenamefont {Hu}\ \emph {et~al.}(2023)\citenamefont {Hu}, \citenamefont {Zhang}, \citenamefont {Yue}, \citenamefont {Liao}, \citenamefont {Liu}, \citenamefont {Zhang}, \citenamefont {Cheng}, \citenamefont {Liu},\ and\ \citenamefont {Christensen}}]{PhysRevLett.131.066601}%
  \BibitemOpen
  \bibfield  {author} {\bibinfo {author} {\bibfnamefont {B.}~\bibnamefont {Hu}}, \bibinfo {author} {\bibfnamefont {Z.}~\bibnamefont {Zhang}}, \bibinfo {author} {\bibfnamefont {Z.}~\bibnamefont {Yue}}, \bibinfo {author} {\bibfnamefont {D.}~\bibnamefont {Liao}}, \bibinfo {author} {\bibfnamefont {Y.}~\bibnamefont {Liu}}, \bibinfo {author} {\bibfnamefont {H.}~\bibnamefont {Zhang}}, \bibinfo {author} {\bibfnamefont {Y.}~\bibnamefont {Cheng}}, \bibinfo {author} {\bibfnamefont {X.}~\bibnamefont {Liu}},\ and\ \bibinfo {author} {\bibfnamefont {J.}~\bibnamefont {Christensen}},\ }\bibfield  {title} {\bibinfo {title} {{Anti-Parity-Time Symmetry in a Su-Schrieffer-Heeger Sonic Lattice}},\ }\href {https://doi.org/10.1103/PhysRevLett.131.066601} {\bibfield  {journal} {\bibinfo  {journal} {Phys. Rev. Lett.}\ }\textbf {\bibinfo {volume} {131}},\ \bibinfo {pages} {066601} (\bibinfo {year} {2023})}\BibitemShut {NoStop}%
\bibitem [{\citenamefont {Hu}\ \emph {et~al.}(2024)\citenamefont {Hu}, \citenamefont {Zhang}, \citenamefont {Liu}, \citenamefont {Liao}, \citenamefont {Zhu}, \citenamefont {Zhang}, \citenamefont {Cheng}, \citenamefont {Liu},\ and\ \citenamefont {Christensen}}]{hu2024engineering}%
  \BibitemOpen
  \bibfield  {author} {\bibinfo {author} {\bibfnamefont {B.}~\bibnamefont {Hu}}, \bibinfo {author} {\bibfnamefont {Z.}~\bibnamefont {Zhang}}, \bibinfo {author} {\bibfnamefont {Y.}~\bibnamefont {Liu}}, \bibinfo {author} {\bibfnamefont {D.}~\bibnamefont {Liao}}, \bibinfo {author} {\bibfnamefont {Y.}~\bibnamefont {Zhu}}, \bibinfo {author} {\bibfnamefont {H.}~\bibnamefont {Zhang}}, \bibinfo {author} {\bibfnamefont {Y.}~\bibnamefont {Cheng}}, \bibinfo {author} {\bibfnamefont {X.}~\bibnamefont {Liu}},\ and\ \bibinfo {author} {\bibfnamefont {J.}~\bibnamefont {Christensen}},\ }\bibfield  {title} {\bibinfo {title} {{Engineering Higher-Order Topological Confinement via Acoustic Non-Hermitian Textures}},\ }\href {https://doi.org/10.1002/adma.202406567} {\bibfield  {journal} {\bibinfo  {journal} {Adv. Mater.}\ }\textbf {\bibinfo {volume} {36}},\ \bibinfo {pages} {2406567} (\bibinfo {year} {2024})}\BibitemShut {NoStop}%
\bibitem [{\citenamefont {Huang}\ \emph {et~al.}(2024)\citenamefont {Huang}, \citenamefont {Huang}, \citenamefont {Shen}, \citenamefont {Yves}, \citenamefont {Pilipchuk}, \citenamefont {Ni}, \citenamefont {Kim}, \citenamefont {Chiang}, \citenamefont {Powell}, \citenamefont {Zhu} \emph {et~al.}}]{huang2024acoustic}%
  \BibitemOpen
  \bibfield  {author} {\bibinfo {author} {\bibfnamefont {L.}~\bibnamefont {Huang}}, \bibinfo {author} {\bibfnamefont {S.}~\bibnamefont {Huang}}, \bibinfo {author} {\bibfnamefont {C.}~\bibnamefont {Shen}}, \bibinfo {author} {\bibfnamefont {S.}~\bibnamefont {Yves}}, \bibinfo {author} {\bibfnamefont {A.~S.}\ \bibnamefont {Pilipchuk}}, \bibinfo {author} {\bibfnamefont {X.}~\bibnamefont {Ni}}, \bibinfo {author} {\bibfnamefont {S.}~\bibnamefont {Kim}}, \bibinfo {author} {\bibfnamefont {Y.~K.}\ \bibnamefont {Chiang}}, \bibinfo {author} {\bibfnamefont {D.~A.}\ \bibnamefont {Powell}}, \bibinfo {author} {\bibfnamefont {J.}~\bibnamefont {Zhu}}, \emph {et~al.},\ }\bibfield  {title} {\bibinfo {title} {{Acoustic resonances in non-Hermitian open systems}},\ }\href {https://doi.org/10.1038/s42254-023-00659-z} {\bibfield  {journal} {\bibinfo  {journal} {Nat. Rev. Phys.}\ }\textbf {\bibinfo {volume} {6}},\ \bibinfo {pages} {11} (\bibinfo {year} {2024})}\BibitemShut {NoStop}%
\bibitem [{\citenamefont {Ghaemi-Dizicheh}(2023)}]{PhysRevB.107.125155}%
  \BibitemOpen
  \bibfield  {author} {\bibinfo {author} {\bibfnamefont {H.}~\bibnamefont {Ghaemi-Dizicheh}},\ }\bibfield  {title} {\bibinfo {title} {Transport effects in non-hermitian nonreciprocal systems: General approach},\ }\href {https://doi.org/10.1103/PhysRevB.107.125155} {\bibfield  {journal} {\bibinfo  {journal} {Phys. Rev. B}\ }\textbf {\bibinfo {volume} {107}},\ \bibinfo {pages} {125155} (\bibinfo {year} {2023})}\BibitemShut {NoStop}%
\bibitem [{\citenamefont {Yan}\ \emph {et~al.}(2024)\citenamefont {Yan}, \citenamefont {Li}, \citenamefont {Sun},\ and\ \citenamefont {Xie}}]{PhysRevB.110.045138}%
  \BibitemOpen
  \bibfield  {author} {\bibinfo {author} {\bibfnamefont {Q.}~\bibnamefont {Yan}}, \bibinfo {author} {\bibfnamefont {H.}~\bibnamefont {Li}}, \bibinfo {author} {\bibfnamefont {Q.-F.}\ \bibnamefont {Sun}},\ and\ \bibinfo {author} {\bibfnamefont {X.~C.}\ \bibnamefont {Xie}},\ }\bibfield  {title} {\bibinfo {title} {Transport theory in non-hermitian systems},\ }\href {https://doi.org/10.1103/PhysRevB.110.045138} {\bibfield  {journal} {\bibinfo  {journal} {Phys. Rev. B}\ }\textbf {\bibinfo {volume} {110}},\ \bibinfo {pages} {045138} (\bibinfo {year} {2024})}\BibitemShut {NoStop}%
\bibitem [{\citenamefont {Li}\ \emph {et~al.}(2025{\natexlab{b}})\citenamefont {Li}, \citenamefont {Chen},\ and\ \citenamefont {Wang}}]{z9m1-3mwb}%
  \BibitemOpen
  \bibfield  {author} {\bibinfo {author} {\bibfnamefont {B.}~\bibnamefont {Li}}, \bibinfo {author} {\bibfnamefont {C.}~\bibnamefont {Chen}},\ and\ \bibinfo {author} {\bibfnamefont {Z.}~\bibnamefont {Wang}},\ }\bibfield  {title} {\bibinfo {title} {Universal non-hermitian transport in disordered systems},\ }\href {https://doi.org/10.1103/z9m1-3mwb} {\bibfield  {journal} {\bibinfo  {journal} {Phys. Rev. Lett.}\ }\textbf {\bibinfo {volume} {135}},\ \bibinfo {pages} {033802} (\bibinfo {year} {2025}{\natexlab{b}})}\BibitemShut {NoStop}%
\bibitem [{\citenamefont {Yang}\ \emph {et~al.}(2020)\citenamefont {Yang}, \citenamefont {Zhang}, \citenamefont {Fang},\ and\ \citenamefont {Hu}}]{PhysRevLett.125.226402}%
  \BibitemOpen
  \bibfield  {author} {\bibinfo {author} {\bibfnamefont {Z.}~\bibnamefont {Yang}}, \bibinfo {author} {\bibfnamefont {K.}~\bibnamefont {Zhang}}, \bibinfo {author} {\bibfnamefont {C.}~\bibnamefont {Fang}},\ and\ \bibinfo {author} {\bibfnamefont {J.}~\bibnamefont {Hu}},\ }\bibfield  {title} {\bibinfo {title} {{Non-Hermitian Bulk-Boundary Correspondence and Auxiliary Generalized Brillouin Zone Theory}},\ }\href {https://doi.org/10.1103/PhysRevLett.125.226402} {\bibfield  {journal} {\bibinfo  {journal} {Phys. Rev. Lett.}\ }\textbf {\bibinfo {volume} {125}},\ \bibinfo {pages} {226402} (\bibinfo {year} {2020})}\BibitemShut {NoStop}%
\bibitem [{\citenamefont {Xiao}\ \emph {et~al.}(2020)\citenamefont {Xiao}, \citenamefont {Deng}, \citenamefont {Wang}, \citenamefont {Zhu}, \citenamefont {Wang}, \citenamefont {Yi},\ and\ \citenamefont {Xue}}]{xiao2020non}%
  \BibitemOpen
  \bibfield  {author} {\bibinfo {author} {\bibfnamefont {L.}~\bibnamefont {Xiao}}, \bibinfo {author} {\bibfnamefont {T.}~\bibnamefont {Deng}}, \bibinfo {author} {\bibfnamefont {K.}~\bibnamefont {Wang}}, \bibinfo {author} {\bibfnamefont {G.}~\bibnamefont {Zhu}}, \bibinfo {author} {\bibfnamefont {Z.}~\bibnamefont {Wang}}, \bibinfo {author} {\bibfnamefont {W.}~\bibnamefont {Yi}},\ and\ \bibinfo {author} {\bibfnamefont {P.}~\bibnamefont {Xue}},\ }\bibfield  {title} {\bibinfo {title} {{Non-Hermitian bulk--boundary correspondence in quantum dynamics}},\ }\href {https://doi.org/10.1038/s41567-020-0836-6} {\bibfield  {journal} {\bibinfo  {journal} {Nat. Phys.}\ }\textbf {\bibinfo {volume} {16}},\ \bibinfo {pages} {761} (\bibinfo {year} {2020})}\BibitemShut {NoStop}%
\bibitem [{\citenamefont {Helbig}\ \emph {et~al.}(2020)\citenamefont {Helbig}, \citenamefont {Hofmann}, \citenamefont {Imhof}, \citenamefont {Abdelghany}, \citenamefont {Kiessling}, \citenamefont {Molenkamp}, \citenamefont {Lee}, \citenamefont {Szameit}, \citenamefont {Greiter},\ and\ \citenamefont {Thomale}}]{helbig2020generalized}%
  \BibitemOpen
  \bibfield  {author} {\bibinfo {author} {\bibfnamefont {T.}~\bibnamefont {Helbig}}, \bibinfo {author} {\bibfnamefont {T.}~\bibnamefont {Hofmann}}, \bibinfo {author} {\bibfnamefont {S.}~\bibnamefont {Imhof}}, \bibinfo {author} {\bibfnamefont {M.}~\bibnamefont {Abdelghany}}, \bibinfo {author} {\bibfnamefont {T.}~\bibnamefont {Kiessling}}, \bibinfo {author} {\bibfnamefont {L.}~\bibnamefont {Molenkamp}}, \bibinfo {author} {\bibfnamefont {C.}~\bibnamefont {Lee}}, \bibinfo {author} {\bibfnamefont {A.}~\bibnamefont {Szameit}}, \bibinfo {author} {\bibfnamefont {M.}~\bibnamefont {Greiter}},\ and\ \bibinfo {author} {\bibfnamefont {R.}~\bibnamefont {Thomale}},\ }\bibfield  {title} {\bibinfo {title} {{Generalized bulk--boundary correspondence in non-Hermitian topolectrical circuits}},\ }\href {https://doi.org/10.1038/s41567-020-0922-9} {\bibfield  {journal} {\bibinfo  {journal} {Nat. Phys.}\ }\textbf {\bibinfo {volume} {16}},\ \bibinfo {pages} {747} (\bibinfo {year} {2020})}\BibitemShut {NoStop}%
\bibitem [{\citenamefont {Cao}\ \emph {et~al.}(2021)\citenamefont {Cao}, \citenamefont {Li},\ and\ \citenamefont {Yang}}]{PhysRevB.103.075126}%
  \BibitemOpen
  \bibfield  {author} {\bibinfo {author} {\bibfnamefont {Y.}~\bibnamefont {Cao}}, \bibinfo {author} {\bibfnamefont {Y.}~\bibnamefont {Li}},\ and\ \bibinfo {author} {\bibfnamefont {X.}~\bibnamefont {Yang}},\ }\bibfield  {title} {\bibinfo {title} {{Non-Hermitian bulk-boundary correspondence in a periodically driven system}},\ }\href {https://doi.org/10.1103/PhysRevB.103.075126} {\bibfield  {journal} {\bibinfo  {journal} {Phys. Rev. B}\ }\textbf {\bibinfo {volume} {103}},\ \bibinfo {pages} {075126} (\bibinfo {year} {2021})}\BibitemShut {NoStop}%
\bibitem [{\citenamefont {Xiao}\ \emph {et~al.}(2024)\citenamefont {Xiao}, \citenamefont {Zhang},\ and\ \citenamefont {Chan}}]{xiao2024restoration}%
  \BibitemOpen
  \bibfield  {author} {\bibinfo {author} {\bibfnamefont {Y.-X.}\ \bibnamefont {Xiao}}, \bibinfo {author} {\bibfnamefont {Z.-Q.}\ \bibnamefont {Zhang}},\ and\ \bibinfo {author} {\bibfnamefont {C.}~\bibnamefont {Chan}},\ }\bibfield  {title} {\bibinfo {title} {{Restoration of non-Hermitian bulk-boundary correspondence by counterbalancing skin effect}},\ }\href {https://doi.org/10.1038/s42005-024-01625-6} {\bibfield  {journal} {\bibinfo  {journal} {Commun. Phys.}\ }\textbf {\bibinfo {volume} {7}},\ \bibinfo {pages} {140} (\bibinfo {year} {2024})}\BibitemShut {NoStop}%
\bibitem [{\citenamefont {Ji}\ and\ \citenamefont {Yang}(2024)}]{Ji_2024}%
  \BibitemOpen
  \bibfield  {author} {\bibinfo {author} {\bibfnamefont {X.}~\bibnamefont {Ji}}\ and\ \bibinfo {author} {\bibfnamefont {X.}~\bibnamefont {Yang}},\ }\bibfield  {title} {\bibinfo {title} {{Generalized bulk-boundary correspondence in periodically driven non-Hermitian systems}},\ }\href {https://doi.org/10.1088/1361-648X/ad2c73} {\bibfield  {journal} {\bibinfo  {journal} {J. Phys.: Condens. Matter}\ }\textbf {\bibinfo {volume} {36}},\ \bibinfo {pages} {243001} (\bibinfo {year} {2024})}\BibitemShut {NoStop}%
\bibitem [{\citenamefont {Joannopoulos}\ \emph {et~al.}(1997)\citenamefont {Joannopoulos}, \citenamefont {Villeneuve},\ and\ \citenamefont {Fan}}]{JOANNOPOULOS1997165}%
  \BibitemOpen
  \bibfield  {author} {\bibinfo {author} {\bibfnamefont {J.}~\bibnamefont {Joannopoulos}}, \bibinfo {author} {\bibfnamefont {P.~R.}\ \bibnamefont {Villeneuve}},\ and\ \bibinfo {author} {\bibfnamefont {S.}~\bibnamefont {Fan}},\ }\bibfield  {title} {\bibinfo {title} {Photonic crystals},\ }\href {https://doi.org/https://doi.org/10.1016/S0038-1098(96)00716-8} {\bibfield  {journal} {\bibinfo  {journal} {Solid State Commun.}\ }\textbf {\bibinfo {volume} {102}},\ \bibinfo {pages} {165} (\bibinfo {year} {1997})},\ \bibinfo {note} {highlights in Condensed Matter Physics and Materials Science}\BibitemShut {NoStop}%
\bibitem [{\citenamefont {Akahane}\ \emph {et~al.}(2003)\citenamefont {Akahane}, \citenamefont {Asano}, \citenamefont {Song},\ and\ \citenamefont {Noda}}]{Akahane2003}%
  \BibitemOpen
  \bibfield  {author} {\bibinfo {author} {\bibfnamefont {Y.}~\bibnamefont {Akahane}}, \bibinfo {author} {\bibfnamefont {T.}~\bibnamefont {Asano}}, \bibinfo {author} {\bibfnamefont {B.-S.}\ \bibnamefont {Song}},\ and\ \bibinfo {author} {\bibfnamefont {S.}~\bibnamefont {Noda}},\ }\bibfield  {title} {\bibinfo {title} {High-{Q} photonic nanocavity in a two-dimensional photonic crystal},\ }\href {https://doi.org/10.1038/nature02063} {\bibfield  {journal} {\bibinfo  {journal} {Nature}\ }\textbf {\bibinfo {volume} {425}},\ \bibinfo {pages} {944} (\bibinfo {year} {2003})}\BibitemShut {NoStop}%
\bibitem [{\citenamefont {Cubukcu}\ \emph {et~al.}(2003)\citenamefont {Cubukcu}, \citenamefont {Aydin}, \citenamefont {Ozbay}, \citenamefont {Foteinopoulou},\ and\ \citenamefont {Soukoulis}}]{Cubukcu2003}%
  \BibitemOpen
  \bibfield  {author} {\bibinfo {author} {\bibfnamefont {E.}~\bibnamefont {Cubukcu}}, \bibinfo {author} {\bibfnamefont {K.}~\bibnamefont {Aydin}}, \bibinfo {author} {\bibfnamefont {E.}~\bibnamefont {Ozbay}}, \bibinfo {author} {\bibfnamefont {S.}~\bibnamefont {Foteinopoulou}},\ and\ \bibinfo {author} {\bibfnamefont {C.~M.}\ \bibnamefont {Soukoulis}},\ }\bibfield  {title} {\bibinfo {title} {Negative refraction by photonic crystals},\ }\href {https://doi.org/10.1038/423604b} {\bibfield  {journal} {\bibinfo  {journal} {Nature}\ }\textbf {\bibinfo {volume} {423}},\ \bibinfo {pages} {604} (\bibinfo {year} {2003})}\BibitemShut {NoStop}%
\bibitem [{\citenamefont {He}\ \emph {et~al.}(2020)\citenamefont {He}, \citenamefont {Addison}, \citenamefont {Mele},\ and\ \citenamefont {Zhen}}]{he2020quadrupole}%
  \BibitemOpen
  \bibfield  {author} {\bibinfo {author} {\bibfnamefont {L.}~\bibnamefont {He}}, \bibinfo {author} {\bibfnamefont {Z.}~\bibnamefont {Addison}}, \bibinfo {author} {\bibfnamefont {E.~J.}\ \bibnamefont {Mele}},\ and\ \bibinfo {author} {\bibfnamefont {B.}~\bibnamefont {Zhen}},\ }\bibfield  {title} {\bibinfo {title} {Quadrupole topological photonic crystals},\ }\href {https://doi.org/10.1038/s41467-020-16916-z} {\bibfield  {journal} {\bibinfo  {journal} {Nat. Commun.}\ }\textbf {\bibinfo {volume} {11}},\ \bibinfo {pages} {3119} (\bibinfo {year} {2020})}\BibitemShut {NoStop}%
\bibitem [{\citenamefont {Ma}\ \emph {et~al.}(2021)\citenamefont {Ma}, \citenamefont {Liu}, \citenamefont {Kudyshev}, \citenamefont {Boltasseva}, \citenamefont {Cai},\ and\ \citenamefont {Liu}}]{ma2021deep}%
  \BibitemOpen
  \bibfield  {author} {\bibinfo {author} {\bibfnamefont {W.}~\bibnamefont {Ma}}, \bibinfo {author} {\bibfnamefont {Z.}~\bibnamefont {Liu}}, \bibinfo {author} {\bibfnamefont {Z.~A.}\ \bibnamefont {Kudyshev}}, \bibinfo {author} {\bibfnamefont {A.}~\bibnamefont {Boltasseva}}, \bibinfo {author} {\bibfnamefont {W.}~\bibnamefont {Cai}},\ and\ \bibinfo {author} {\bibfnamefont {Y.}~\bibnamefont {Liu}},\ }\bibfield  {title} {\bibinfo {title} {Deep learning for the design of photonic structures},\ }\href {https://doi.org/10.1038/s41566-020-0685-y} {\bibfield  {journal} {\bibinfo  {journal} {Nat. Photonics}\ }\textbf {\bibinfo {volume} {15}},\ \bibinfo {pages} {77} (\bibinfo {year} {2021})}\BibitemShut {NoStop}%
\bibitem [{\citenamefont {Ota}\ \emph {et~al.}(2018)\citenamefont {Ota}, \citenamefont {Katsumi}, \citenamefont {Watanabe}, \citenamefont {Iwamoto},\ and\ \citenamefont {Arakawa}}]{ota2018topological}%
  \BibitemOpen
  \bibfield  {author} {\bibinfo {author} {\bibfnamefont {Y.}~\bibnamefont {Ota}}, \bibinfo {author} {\bibfnamefont {R.}~\bibnamefont {Katsumi}}, \bibinfo {author} {\bibfnamefont {K.}~\bibnamefont {Watanabe}}, \bibinfo {author} {\bibfnamefont {S.}~\bibnamefont {Iwamoto}},\ and\ \bibinfo {author} {\bibfnamefont {Y.}~\bibnamefont {Arakawa}},\ }\bibfield  {title} {\bibinfo {title} {Topological photonic crystal nanocavity laser},\ }\href {https://doi.org/10.1038/s42005-018-0083-7} {\bibfield  {journal} {\bibinfo  {journal} {Commun. Phys.}\ }\textbf {\bibinfo {volume} {1}},\ \bibinfo {pages} {86} (\bibinfo {year} {2018})}\BibitemShut {NoStop}%
\bibitem [{\citenamefont {Bandres}\ \emph {et~al.}(2018)\citenamefont {Bandres}, \citenamefont {Wittek}, \citenamefont {Harari}, \citenamefont {Parto}, \citenamefont {Ren}, \citenamefont {Segev}, \citenamefont {Christodoulides},\ and\ \citenamefont {Khajavikhan}}]{doi:10.1126/science.aar4005}%
  \BibitemOpen
  \bibfield  {author} {\bibinfo {author} {\bibfnamefont {M.~A.}\ \bibnamefont {Bandres}}, \bibinfo {author} {\bibfnamefont {S.}~\bibnamefont {Wittek}}, \bibinfo {author} {\bibfnamefont {G.}~\bibnamefont {Harari}}, \bibinfo {author} {\bibfnamefont {M.}~\bibnamefont {Parto}}, \bibinfo {author} {\bibfnamefont {J.}~\bibnamefont {Ren}}, \bibinfo {author} {\bibfnamefont {M.}~\bibnamefont {Segev}}, \bibinfo {author} {\bibfnamefont {D.~N.}\ \bibnamefont {Christodoulides}},\ and\ \bibinfo {author} {\bibfnamefont {M.}~\bibnamefont {Khajavikhan}},\ }\bibfield  {title} {\bibinfo {title} {Topological insulator laser: Experiments},\ }\href {https://doi.org/10.1126/science.aar4005} {\bibfield  {journal} {\bibinfo  {journal} {Science}\ }\textbf {\bibinfo {volume} {359}},\ \bibinfo {pages} {eaar4005} (\bibinfo {year} {2018})}\BibitemShut {NoStop}%
\bibitem [{\citenamefont {Yang}\ \emph {et~al.}(2019)\citenamefont {Yang}, \citenamefont {Gao}, \citenamefont {Xue}, \citenamefont {Zhang}, \citenamefont {He}, \citenamefont {Yang}, \citenamefont {Singh}, \citenamefont {Chong}, \citenamefont {Zhang},\ and\ \citenamefont {Chen}}]{yang2019realization}%
  \BibitemOpen
  \bibfield  {author} {\bibinfo {author} {\bibfnamefont {Y.}~\bibnamefont {Yang}}, \bibinfo {author} {\bibfnamefont {Z.}~\bibnamefont {Gao}}, \bibinfo {author} {\bibfnamefont {H.}~\bibnamefont {Xue}}, \bibinfo {author} {\bibfnamefont {L.}~\bibnamefont {Zhang}}, \bibinfo {author} {\bibfnamefont {M.}~\bibnamefont {He}}, \bibinfo {author} {\bibfnamefont {Z.}~\bibnamefont {Yang}}, \bibinfo {author} {\bibfnamefont {R.}~\bibnamefont {Singh}}, \bibinfo {author} {\bibfnamefont {Y.}~\bibnamefont {Chong}}, \bibinfo {author} {\bibfnamefont {B.}~\bibnamefont {Zhang}},\ and\ \bibinfo {author} {\bibfnamefont {H.}~\bibnamefont {Chen}},\ }\bibfield  {title} {\bibinfo {title} {Realization of a three-dimensional photonic topological insulator},\ }\href {https://doi.org/10.1038/s41586-018-0829-0} {\bibfield  {journal} {\bibinfo  {journal} {Nature}\ }\textbf {\bibinfo {volume} {565}},\ \bibinfo {pages} {622} (\bibinfo {year} {2019})}\BibitemShut {NoStop}%
\bibitem [{\citenamefont {Chen}\ \emph {et~al.}(2019)\citenamefont {Chen}, \citenamefont {Deng}, \citenamefont {Shi}, \citenamefont {Zhao}, \citenamefont {Chen},\ and\ \citenamefont {Dong}}]{PhysRevLett.122.233902}%
  \BibitemOpen
  \bibfield  {author} {\bibinfo {author} {\bibfnamefont {X.-D.}\ \bibnamefont {Chen}}, \bibinfo {author} {\bibfnamefont {W.-M.}\ \bibnamefont {Deng}}, \bibinfo {author} {\bibfnamefont {F.-L.}\ \bibnamefont {Shi}}, \bibinfo {author} {\bibfnamefont {F.-L.}\ \bibnamefont {Zhao}}, \bibinfo {author} {\bibfnamefont {M.}~\bibnamefont {Chen}},\ and\ \bibinfo {author} {\bibfnamefont {J.-W.}\ \bibnamefont {Dong}},\ }\bibfield  {title} {\bibinfo {title} {Direct observation of corner states in second-order topological photonic crystal slabs},\ }\href {https://doi.org/10.1103/PhysRevLett.122.233902} {\bibfield  {journal} {\bibinfo  {journal} {Phys. Rev. Lett.}\ }\textbf {\bibinfo {volume} {122}},\ \bibinfo {pages} {233902} (\bibinfo {year} {2019})}\BibitemShut {NoStop}%
\bibitem [{\citenamefont {Lu}\ \emph {et~al.}(2014)\citenamefont {Lu}, \citenamefont {Joannopoulos},\ and\ \citenamefont {Solja{\v{c}}i{\'c}}}]{lu2014topological}%
  \BibitemOpen
  \bibfield  {author} {\bibinfo {author} {\bibfnamefont {L.}~\bibnamefont {Lu}}, \bibinfo {author} {\bibfnamefont {J.~D.}\ \bibnamefont {Joannopoulos}},\ and\ \bibinfo {author} {\bibfnamefont {M.}~\bibnamefont {Solja{\v{c}}i{\'c}}},\ }\bibfield  {title} {\bibinfo {title} {{Topological photonics}},\ }\href {https://doi.org/10.1038/nphoton.2014.248} {\bibfield  {journal} {\bibinfo  {journal} {Nat. Photonics}\ }\textbf {\bibinfo {volume} {8}},\ \bibinfo {pages} {821} (\bibinfo {year} {2014})}\BibitemShut {NoStop}%
\bibitem [{\citenamefont {Khanikaev}\ and\ \citenamefont {Shvets}(2017)}]{khanikaev2017two}%
  \BibitemOpen
  \bibfield  {author} {\bibinfo {author} {\bibfnamefont {A.~B.}\ \bibnamefont {Khanikaev}}\ and\ \bibinfo {author} {\bibfnamefont {G.}~\bibnamefont {Shvets}},\ }\bibfield  {title} {\bibinfo {title} {{Two-dimensional topological photonics}},\ }\href {https://doi.org/10.1038/s41566-017-0048-5} {\bibfield  {journal} {\bibinfo  {journal} {Nat. Photonics}\ }\textbf {\bibinfo {volume} {11}},\ \bibinfo {pages} {763} (\bibinfo {year} {2017})}\BibitemShut {NoStop}%
\bibitem [{\citenamefont {Xie}\ \emph {et~al.}(2018)\citenamefont {Xie}, \citenamefont {Wang}, \citenamefont {Zhu}, \citenamefont {Lu}, \citenamefont {Wang},\ and\ \citenamefont {Chen}}]{Xie:18}%
  \BibitemOpen
  \bibfield  {author} {\bibinfo {author} {\bibfnamefont {B.-Y.}\ \bibnamefont {Xie}}, \bibinfo {author} {\bibfnamefont {H.-F.}\ \bibnamefont {Wang}}, \bibinfo {author} {\bibfnamefont {X.-Y.}\ \bibnamefont {Zhu}}, \bibinfo {author} {\bibfnamefont {M.-H.}\ \bibnamefont {Lu}}, \bibinfo {author} {\bibfnamefont {Z.~D.}\ \bibnamefont {Wang}},\ and\ \bibinfo {author} {\bibfnamefont {Y.-F.}\ \bibnamefont {Chen}},\ }\bibfield  {title} {\bibinfo {title} {{Photonics meets topology}},\ }\href {https://doi.org/10.1364/OE.26.024531} {\bibfield  {journal} {\bibinfo  {journal} {Opt. Express}\ }\textbf {\bibinfo {volume} {26}},\ \bibinfo {pages} {24531} (\bibinfo {year} {2018})}\BibitemShut {NoStop}%
\bibitem [{\citenamefont {Ozawa}\ \emph {et~al.}(2019)\citenamefont {Ozawa}, \citenamefont {Price}, \citenamefont {Amo}, \citenamefont {Goldman}, \citenamefont {Hafezi}, \citenamefont {Lu}, \citenamefont {Rechtsman}, \citenamefont {Schuster}, \citenamefont {Simon}, \citenamefont {Zilberberg},\ and\ \citenamefont {Carusotto}}]{RevModPhys.91.015006}%
  \BibitemOpen
  \bibfield  {author} {\bibinfo {author} {\bibfnamefont {T.}~\bibnamefont {Ozawa}}, \bibinfo {author} {\bibfnamefont {H.~M.}\ \bibnamefont {Price}}, \bibinfo {author} {\bibfnamefont {A.}~\bibnamefont {Amo}}, \bibinfo {author} {\bibfnamefont {N.}~\bibnamefont {Goldman}}, \bibinfo {author} {\bibfnamefont {M.}~\bibnamefont {Hafezi}}, \bibinfo {author} {\bibfnamefont {L.}~\bibnamefont {Lu}}, \bibinfo {author} {\bibfnamefont {M.~C.}\ \bibnamefont {Rechtsman}}, \bibinfo {author} {\bibfnamefont {D.}~\bibnamefont {Schuster}}, \bibinfo {author} {\bibfnamefont {J.}~\bibnamefont {Simon}}, \bibinfo {author} {\bibfnamefont {O.}~\bibnamefont {Zilberberg}},\ and\ \bibinfo {author} {\bibfnamefont {I.}~\bibnamefont {Carusotto}},\ }\bibfield  {title} {\bibinfo {title} {{Topological photonics}},\ }\href {https://doi.org/10.1103/RevModPhys.91.015006} {\bibfield  {journal} {\bibinfo  {journal} {Rev. Mod. Phys.}\ }\textbf {\bibinfo {volume} {91}},\ \bibinfo {pages} {015006} (\bibinfo {year} {2019})}\BibitemShut {NoStop}%
\bibitem [{\citenamefont {Ota}\ \emph {et~al.}(2020)\citenamefont {Ota}, \citenamefont {Takata}, \citenamefont {Ozawa}, \citenamefont {Amo}, \citenamefont {Jia}, \citenamefont {Kante}, \citenamefont {Notomi}, \citenamefont {Arakawa},\ and\ \citenamefont {Iwamoto}}]{OtaTakataOzawaAmoJiaKanteNotomiArakawaIwamoto+2020+547+567}%
  \BibitemOpen
  \bibfield  {author} {\bibinfo {author} {\bibfnamefont {Y.}~\bibnamefont {Ota}}, \bibinfo {author} {\bibfnamefont {K.}~\bibnamefont {Takata}}, \bibinfo {author} {\bibfnamefont {T.}~\bibnamefont {Ozawa}}, \bibinfo {author} {\bibfnamefont {A.}~\bibnamefont {Amo}}, \bibinfo {author} {\bibfnamefont {Z.}~\bibnamefont {Jia}}, \bibinfo {author} {\bibfnamefont {B.}~\bibnamefont {Kante}}, \bibinfo {author} {\bibfnamefont {M.}~\bibnamefont {Notomi}}, \bibinfo {author} {\bibfnamefont {Y.}~\bibnamefont {Arakawa}},\ and\ \bibinfo {author} {\bibfnamefont {S.}~\bibnamefont {Iwamoto}},\ }\bibfield  {title} {\bibinfo {title} {{Active topological photonics}},\ }\href {https://doi.org/doi:10.1515/nanoph-2019-0376} {\bibfield  {journal} {\bibinfo  {journal} {Nanophotonics}\ }\textbf {\bibinfo {volume} {9}},\ \bibinfo {pages} {547} (\bibinfo {year} {2020})}\BibitemShut {NoStop}%
\bibitem [{\citenamefont {Kim}\ \emph {et~al.}(2020)\citenamefont {Kim}, \citenamefont {Jacob},\ and\ \citenamefont {Rho}}]{kim2020recent}%
  \BibitemOpen
  \bibfield  {author} {\bibinfo {author} {\bibfnamefont {M.}~\bibnamefont {Kim}}, \bibinfo {author} {\bibfnamefont {Z.}~\bibnamefont {Jacob}},\ and\ \bibinfo {author} {\bibfnamefont {J.}~\bibnamefont {Rho}},\ }\bibfield  {title} {\bibinfo {title} {{Recent advances in 2D, 3D and higher-order topological photonics}},\ }\href {https://doi.org/10.1038/s41377-020-0331-y} {\bibfield  {journal} {\bibinfo  {journal} {Light Sci. Appl.}\ }\textbf {\bibinfo {volume} {9}},\ \bibinfo {pages} {130} (\bibinfo {year} {2020})}\BibitemShut {NoStop}%
\bibitem [{\citenamefont {Xue}\ \emph {et~al.}(2021)\citenamefont {Xue}, \citenamefont {Yang},\ and\ \citenamefont {Zhang}}]{xue2021topological}%
  \BibitemOpen
  \bibfield  {author} {\bibinfo {author} {\bibfnamefont {H.}~\bibnamefont {Xue}}, \bibinfo {author} {\bibfnamefont {Y.}~\bibnamefont {Yang}},\ and\ \bibinfo {author} {\bibfnamefont {B.}~\bibnamefont {Zhang}},\ }\bibfield  {title} {\bibinfo {title} {{Topological valley photonics: physics and device applications}},\ }\href {https://doi.org/10.1002/adpr.202100013} {\bibfield  {journal} {\bibinfo  {journal} {Adv. Photonics Res.}\ }\textbf {\bibinfo {volume} {2}},\ \bibinfo {pages} {2100013} (\bibinfo {year} {2021})}\BibitemShut {NoStop}%
\bibitem [{\citenamefont {Xie}\ \emph {et~al.}(2023)\citenamefont {Xie}, \citenamefont {Jin},\ and\ \citenamefont {Song}}]{XIE2023255}%
  \BibitemOpen
  \bibfield  {author} {\bibinfo {author} {\bibfnamefont {L.}~\bibnamefont {Xie}}, \bibinfo {author} {\bibfnamefont {L.}~\bibnamefont {Jin}},\ and\ \bibinfo {author} {\bibfnamefont {Z.}~\bibnamefont {Song}},\ }\bibfield  {title} {\bibinfo {title} {{Antihelical edge states in two-dimensional photonic topological metals}},\ }\href {https://doi.org/https://doi.org/10.1016/j.scib.2023.01.018} {\bibfield  {journal} {\bibinfo  {journal} {Sci. Bull.}\ }\textbf {\bibinfo {volume} {68}},\ \bibinfo {pages} {255} (\bibinfo {year} {2023})}\BibitemShut {NoStop}%
\bibitem [{\citenamefont {Zhang}\ \emph {et~al.}(2024{\natexlab{b}})\citenamefont {Zhang}, \citenamefont {Wang}, \citenamefont {Song},\ and\ \citenamefont {Cai}}]{ZHANG2024111353}%
  \BibitemOpen
  \bibfield  {author} {\bibinfo {author} {\bibfnamefont {L.}~\bibnamefont {Zhang}}, \bibinfo {author} {\bibfnamefont {B.}~\bibnamefont {Wang}}, \bibinfo {author} {\bibfnamefont {S.}~\bibnamefont {Song}},\ and\ \bibinfo {author} {\bibfnamefont {J.}~\bibnamefont {Cai}},\ }\bibfield  {title} {\bibinfo {title} {{Edge and corner states in non-Hermitian second-order topological photonic crystals}},\ }\href {https://doi.org/https://doi.org/10.1016/j.optlastec.2024.111353} {\bibfield  {journal} {\bibinfo  {journal} {Opt. Laser Technol.}\ }\textbf {\bibinfo {volume} {179}},\ \bibinfo {pages} {111353} (\bibinfo {year} {2024}{\natexlab{b}})}\BibitemShut {NoStop}%
\bibitem [{\citenamefont {Khalaf}(2018)}]{PhysRevB.97.205136}%
  \BibitemOpen
  \bibfield  {author} {\bibinfo {author} {\bibfnamefont {E.}~\bibnamefont {Khalaf}},\ }\bibfield  {title} {\bibinfo {title} {{Higher-order topological insulators and superconductors protected by inversion symmetry}},\ }\href {https://doi.org/10.1103/PhysRevB.97.205136} {\bibfield  {journal} {\bibinfo  {journal} {Phys. Rev. B}\ }\textbf {\bibinfo {volume} {97}},\ \bibinfo {pages} {205136} (\bibinfo {year} {2018})}\BibitemShut {NoStop}%
\bibitem [{\citenamefont {Zhang}\ \emph {et~al.}(2019{\natexlab{b}})\citenamefont {Zhang}, \citenamefont {Xie}, \citenamefont {Wang}, \citenamefont {Xu}, \citenamefont {Tian}, \citenamefont {Jiang}, \citenamefont {Lu},\ and\ \citenamefont {Chen}}]{zhang2019dimensional}%
  \BibitemOpen
  \bibfield  {author} {\bibinfo {author} {\bibfnamefont {X.}~\bibnamefont {Zhang}}, \bibinfo {author} {\bibfnamefont {B.-Y.}\ \bibnamefont {Xie}}, \bibinfo {author} {\bibfnamefont {H.-F.}\ \bibnamefont {Wang}}, \bibinfo {author} {\bibfnamefont {X.}~\bibnamefont {Xu}}, \bibinfo {author} {\bibfnamefont {Y.}~\bibnamefont {Tian}}, \bibinfo {author} {\bibfnamefont {J.-H.}\ \bibnamefont {Jiang}}, \bibinfo {author} {\bibfnamefont {M.-H.}\ \bibnamefont {Lu}},\ and\ \bibinfo {author} {\bibfnamefont {Y.-F.}\ \bibnamefont {Chen}},\ }\bibfield  {title} {\bibinfo {title} {{Dimensional hierarchy of higher-order topology in three-dimensional sonic crystals}},\ }\href {https://doi.org/10.1038/s41467-019-13333-9} {\bibfield  {journal} {\bibinfo  {journal} {Nat. Commun.}\ }\textbf {\bibinfo {volume} {10}},\ \bibinfo {pages} {5331} (\bibinfo {year} {2019}{\natexlab{b}})}\BibitemShut {NoStop}%
\bibitem [{\citenamefont {Xie}\ \emph {et~al.}(2021)\citenamefont {Xie}, \citenamefont {Wang}, \citenamefont {Zhang}, \citenamefont {Zhan}, \citenamefont {Jiang}, \citenamefont {Lu},\ and\ \citenamefont {Chen}}]{xie2021higher}%
  \BibitemOpen
  \bibfield  {author} {\bibinfo {author} {\bibfnamefont {B.}~\bibnamefont {Xie}}, \bibinfo {author} {\bibfnamefont {H.-X.}\ \bibnamefont {Wang}}, \bibinfo {author} {\bibfnamefont {X.}~\bibnamefont {Zhang}}, \bibinfo {author} {\bibfnamefont {P.}~\bibnamefont {Zhan}}, \bibinfo {author} {\bibfnamefont {J.-H.}\ \bibnamefont {Jiang}}, \bibinfo {author} {\bibfnamefont {M.}~\bibnamefont {Lu}},\ and\ \bibinfo {author} {\bibfnamefont {Y.}~\bibnamefont {Chen}},\ }\bibfield  {title} {\bibinfo {title} {{Higher-order band topology}},\ }\href {https://doi.org/10.1038/s42254-021-00323-4} {\bibfield  {journal} {\bibinfo  {journal} {Nat. Rev. Phys.}\ }\textbf {\bibinfo {volume} {3}},\ \bibinfo {pages} {520} (\bibinfo {year} {2021})}\BibitemShut {NoStop}%
\bibitem [{\citenamefont {Atala}\ \emph {et~al.}(2013)\citenamefont {Atala}, \citenamefont {Aidelsburger}, \citenamefont {Barreiro}, \citenamefont {Abanin}, \citenamefont {Kitagawa}, \citenamefont {Demler},\ and\ \citenamefont {Bloch}}]{atala2013direct}%
  \BibitemOpen
  \bibfield  {author} {\bibinfo {author} {\bibfnamefont {M.}~\bibnamefont {Atala}}, \bibinfo {author} {\bibfnamefont {M.}~\bibnamefont {Aidelsburger}}, \bibinfo {author} {\bibfnamefont {J.~T.}\ \bibnamefont {Barreiro}}, \bibinfo {author} {\bibfnamefont {D.}~\bibnamefont {Abanin}}, \bibinfo {author} {\bibfnamefont {T.}~\bibnamefont {Kitagawa}}, \bibinfo {author} {\bibfnamefont {E.}~\bibnamefont {Demler}},\ and\ \bibinfo {author} {\bibfnamefont {I.}~\bibnamefont {Bloch}},\ }\bibfield  {title} {\bibinfo {title} {{Direct measurement of the Zak phase in topological Bloch bands}},\ }\href {https://doi.org/10.1038/nphys2790} {\bibfield  {journal} {\bibinfo  {journal} {Nat. Phys.}\ }\textbf {\bibinfo {volume} {9}},\ \bibinfo {pages} {795} (\bibinfo {year} {2013})}\BibitemShut {NoStop}%
\bibitem [{\citenamefont {Yuce}\ and\ \citenamefont {Ramezani}(2019)}]{PhysRevA.100.032102}%
  \BibitemOpen
  \bibfield  {author} {\bibinfo {author} {\bibfnamefont {C.}~\bibnamefont {Yuce}}\ and\ \bibinfo {author} {\bibfnamefont {H.}~\bibnamefont {Ramezani}},\ }\bibfield  {title} {\bibinfo {title} {{Topological states in a non-Hermitian two-dimensional Su-Schrieffer-Heeger model}},\ }\href {https://doi.org/10.1103/PhysRevA.100.032102} {\bibfield  {journal} {\bibinfo  {journal} {Phys. Rev. A}\ }\textbf {\bibinfo {volume} {100}},\ \bibinfo {pages} {032102} (\bibinfo {year} {2019})}\BibitemShut {NoStop}%
\bibitem [{\citenamefont {Wang}\ \emph {et~al.}(2019)\citenamefont {Wang}, \citenamefont {Guo},\ and\ \citenamefont {Jiang}}]{Wang_2019}%
  \BibitemOpen
  \bibfield  {author} {\bibinfo {author} {\bibfnamefont {H.-X.}\ \bibnamefont {Wang}}, \bibinfo {author} {\bibfnamefont {G.-Y.}\ \bibnamefont {Guo}},\ and\ \bibinfo {author} {\bibfnamefont {J.-H.}\ \bibnamefont {Jiang}},\ }\bibfield  {title} {\bibinfo {title} {{Band topology in classical waves: Wilson-loop approach to topological numbers and fragile topology}},\ }\href {https://doi.org/10.1088/1367-2630/ab3f71} {\bibfield  {journal} {\bibinfo  {journal} {New J. Phys.}\ }\textbf {\bibinfo {volume} {21}},\ \bibinfo {pages} {093029} (\bibinfo {year} {2019})}\BibitemShut {NoStop}%
\bibitem [{\citenamefont {Chen}\ \emph {et~al.}(2021)\citenamefont {Chen}, \citenamefont {Jiang}, \citenamefont {Zhang}, \citenamefont {Zhao}, \citenamefont {Lan},\ and\ \citenamefont {Sha}}]{PhysRevA.104.033501}%
  \BibitemOpen
  \bibfield  {author} {\bibinfo {author} {\bibfnamefont {M.~L.~N.}\ \bibnamefont {Chen}}, \bibinfo {author} {\bibfnamefont {L.~J.}\ \bibnamefont {Jiang}}, \bibinfo {author} {\bibfnamefont {S.}~\bibnamefont {Zhang}}, \bibinfo {author} {\bibfnamefont {R.}~\bibnamefont {Zhao}}, \bibinfo {author} {\bibfnamefont {Z.}~\bibnamefont {Lan}},\ and\ \bibinfo {author} {\bibfnamefont {W.~E.~I.}\ \bibnamefont {Sha}},\ }\bibfield  {title} {\bibinfo {title} {{Comparative study of Hermitian and non-Hermitian topological dielectric photonic crystals}},\ }\href {https://doi.org/10.1103/PhysRevA.104.033501} {\bibfield  {journal} {\bibinfo  {journal} {Phys. Rev. A}\ }\textbf {\bibinfo {volume} {104}},\ \bibinfo {pages} {033501} (\bibinfo {year} {2021})}\BibitemShut {NoStop}%
\end{thebibliography}%

\end{document}